\documentclass{ws-ijmpa}
\usepackage[super,compress]{cite}
\usepackage{graphicx}
\usepackage{amsbsy}
\usepackage{epstopdf}

\def\mprp{\mbox{\tiny $\bot$}}
\def\mprl{\mbox{\tiny $\|$}}

\def\beq{\begin{eqnarray}}
\def\eeq{\end{eqnarray}}

\newcommand{\bs}{\boldsymbol} 
\newcommand{\ii}{\mathrm{i}} 
\newcommand{\dd}{\mathrm{d}} 
\newcommand{\eee}{\mathrm{e}} 

\begin{document}
\markboth{A.\,V.~Kuznetsov, D.\,A.~Rumyantsev, D.\,M.~Shlenev}
{Generalized two-point tree-level amplitude  \dots}
%
\catchline{}{}{}{}{}
%

\title{
GENERALIZED TWO-POINT TREE-LEVEL AMPLITUDE $jf \to j^{\, \prime} f^{\, \prime}$
IN A MAGNETIZED MEDIUM
}

\author{A.\,V.~KUZNETSOV, D.\,A.~RUMYANTSEV, D.\,M.~SHLENEV}

\address{Division of Theoretical Physics, Department of Physics,\\
Yaroslavl State P.\,G.~Demidov University, Sovietskaya 14,\\
150000 Yaroslavl, Russian Federation\\
avkuzn@uniyar.ac.ru, rda@uniyar.ac.ru, allen\_caleb@rambler.ru}



\maketitle

\begin{history}
\received{Day Month Year}
\revised{Day Month Year}
\end{history}

\begin{abstract}
The tree-level two-point amplitudes for the transitions 
$jf \to j^{\, \prime} f^{\, \prime}$, where $f$ is a fermion and $j$ is a generalized current, 
in a constant uniform magnetic field of an arbitrary strength and in  
charged fermion plasma, for the $jff$ 
interaction vertices of the scalar, pseudoscalar, vector and  
axial-vector types have been calculated. 
The generalized current $j$ could mean the field operator of a boson, or a current 
consisting of fermions, e.g. the neutrino current. 
The particular cases of a very strong magnetic field, 
and of the coherent scattering off the real fermions without change of
their states (the ``forward'' scattering) have been analysed. The contribution of the neutrino photoproduction process, 
$\gamma e\to e \nu \bar \nu$,  
to the neutrino emissivity has been calculated with taking account of a possible resonance on the 
virtual electron. 

\end{abstract}

\ccode{PACS numbers: 12.20.Ds, 14.60.Cd, 97.10.Ld, 94.30.-d}


\section{Introduction}
\label{sec:Introduction}

Nowadays, there exists rather keen interest to astrophysical objects
with the scale of the magnetic field strength near  
the critical value of $B_e = m_e^2/e \simeq 4.41\times 10^{13}$~G~\footnote{We use natural units
$c = \hbar = k_{\rm{B}} = 1$, $m_e$ is the electron mass, and $e$ is the elementary
charge, $m_f$ and $e_f$ are the fermion mass and the fermion
charge.}. 
This group of objects includes the radio pulsars and the so-called
magnetars, which are the neutron stars featuring
the magnetic field strengths from $10^{12}$~G (radio pulsars) to  
$4\times 10^{14}$~G 
(magnetars), 
see Ref.~\refcite{Olausen:2014} and the papers cited therein. 
The spectra analysis of these objects also provides an evidence for the
presence of electron-positron plasma in the radio pulsar and magnetar environment, 
with the minimum magnetospheric plasma density being of the order 
of the Goldreich-Julian density~\cite{GJ:1969}: 
\begin{eqnarray}
\label{eq:ngj}
n_{GJ} \simeq 
 3 \times 10^{13}\, \mbox{cm}^{-3} 
\left (\frac{B}{100B_e} \right )\left (\frac{10\,\mbox{s}}{P} \right ) \, ,
\end{eqnarray}                                         
where $P$ is the rotational period.
It is well-known that strong magnetic field and/or plasma could have an essential 
influence on various quantum 
processes~\cite{Lai:2001,Harding:2006qn,KM_Book_2003,KM_Book_2013}, because the external active medium 
catalyses the processes, by changing their kinematics and inducing new interactions. Therefore,
the effects of magnetized plasma on microscopic physics should be
incorporated in the magnetosphere models of strongly magnetized neutron stars.  
In the present paper we consider the two-point processes, because such reactions 
can have  possible resonant behavior, and therefore they could be 
very interesting for astrophysical applications~\cite{Lyutikov:2006}.  
                                                           
The investigation of the two-point processes in an external active medium (electromagnetic 
field and/or plasma) has a rather long history. 
The most general expression for a
two-vertex loop amplitude of the form $j \to f \bar f \to j^{\, \prime}$
in a pure constant uniform magnetic field and in a crossed
field was obtained previously in Ref.~\refcite{Borovkov:1999}, where all possible combinations
of scalar, pseudoscalar, vector, and axial-vector 
interactions of the generalized currents $j$ and $j^{\, \prime}$
with fermions were considered.
The generalized current $j$ could mean the field operator of a boson, or a current 
consisting of fermions, e.g. the neutrino current. 

The typical example of a tree-level process 
with two vector vertices in the presence of magnetized plasma is the
Compton scattering, $\gamma e \to \gamma e$,
as a possible channel of the radiation spectra formation. 
In this case, both generalized currents $j$ and $j^{\, \prime}$ mean the photon field operators.
This process  was studied in 
a number of papers, see e.g. Refs.~\refcite{Herold:1979}--\refcite{Weise:2014}, 
but the results were presented there in the form without taking 
account of the photon dispersion properties. In Ref.~\refcite{RCh09} 
this neglect was corrected. The expression for the Compton scattering 
amplitude, with the initial and final electrons being on the lowest Landau level 
was presented in Ref.~\refcite{RCh09} in the explicit Lorentz invariant form. 
The other example of the Compton like process with the vector and 
axial-vector vertices, the photon transition into the neutrino pair 
in the presence of magnetized plasma, $\gamma e \to e \nu \bar \nu$, was studied in Ref.~\refcite{Kennett:1998}. 
In this case, in our terms, the initial generalized current $j$ means the photon field operator, while the final generalized 
current $j^{\, \prime}$ means the neutrino current. The local limit of the weak interaction is supposed to be valid. 
One of our goals in this paper is to improve the approach of Ref.~\refcite{Kennett:1998}, 
in order to present the results in a manifestly covariant form. 
Additionally, as we believe, our results would be better applicable for an analysis of the other photon-fermion 
scattering processes with the production of exotic particles, such as axion, neutralino, etc.

Thus, we consider the tree-level two-point amplitude for the transition of the type
$jf \to j^{\, \prime} f^{\prime}$ with the intermediate virtual fermion state. 
The analysis is performed in a constant uniform magnetic field and 
charged fermion plasma, for different combinations of the vertices that were used in Ref.~\refcite{Borovkov:1999}.  Particularly, we generalize the results obtained 
in Ref.~\refcite{Borovkov:1999}, to the case of magnetized plasma, since such a situation looks 
the most realistic for astrophysical objects. Such a generalization was performed in part 
in Ref.~\refcite{Shabad} for the case of the photon polarization operator in a magnetized electron-positron plasma. 
                  
The paper is organized as follows. 
In Sec.~\ref{Sec:2}, we calculate the scattering amplitudes for different spin states  
of the initial and final fermions. We present here only the amplitudes 
for the case when both vertices are of the pseudoscalar type. The total set of the amplitudes
for the $jff$ interaction vertices of the scalar, pseudoscalar, vector and  
axial-vector types, in a constant uniform magnetic field of an arbitrary strength and in  
charged fermion plasma can be found in the extended paper~\cite{Kuznetsov:2013}. 
All the amplitudes are presented in the explicit Lorentz and gauge invariant forms. 
The application of the 
obtained results to the calculation of the neutrino photoproduction process amplitude and 
other characteristics in the resonant case is given in Sec.~\ref{Sec:4a}.   
Final comments and discussion of the obtained results and possible astrophysical applications 
are given in Sec.~\ref{Sec:Discussion}.
In~\ref{Append:A}, we present  the fermion wave functions used in our analysis, namely,
the solutions of the Dirac equation in external magnetic field, 
being the eigenfunctions of the magnetic moment operator.
In the next two Appendices, we present the expressions for the amplitudes in the special cases 
where they can be essentially simplified.  
In~\ref{Append:B}, we consider the particular case, when the initial and final fermions
occupy the ground Landau level (the strong field limit), for all types of the $jff$ interaction vertices. 
A coherent scattering of neutral particles off the real fermions without change of
their states (the ``forward'' scattering) is analysed in~\ref{Append:C}. 


\section{The set of expressions for the amplitudes}
\label{Sec:2}


The generalized amplitude of the transition $jf \to j^{\, \prime} f^{\, \prime}$ 
will be analyzed 
by using the effective Lagrangian for the interaction of a generalized current $j$ 
with fermions in the form
\begin{eqnarray}
{\cal L}(X) \, = \, \sum \limits_{k} g_k 
[\bar \Psi_f (X) \Gamma_k \Psi_f(X)] j_k(X), 
\label{eq:L}
\end{eqnarray}
\noindent
where the generalized index $k = S, P, V, A$ numbers the matrices 
$\Gamma_k$:   
$\Gamma_S = 1, \, \Gamma_P = \gamma_5, \, \Gamma_V = \gamma_{\alpha},
\, \Gamma_A = \gamma_{\alpha} \gamma_5$; 
$j_k(X)$ are the generalized currents ($j_S$, $j_P$, $j_{V\alpha}$ or $j_{A\alpha}$) 
or the field operators of single particles, e.g. of the photon or axion, see below, 
$g_k$ are the corresponding coupling constants, and
$\Psi_f(X)$ are the fermion wave functions.

Indeed, using the Lagrangian~(\ref{eq:L}), one can describe a large class of interactions. 
For example, it may be:

i) the Lagrangian of the electromagnetic interaction, when $k=V$, $g_V = - e_f$, 
$\Gamma_V j_V = \gamma^{\mu} A_\mu$, 
$A^\mu$ is the four-potential of the quantized electromagnetic field:
\begin{eqnarray}
{\cal L}(X) \, = \, - e_f \, [\bar \Psi_f (X) \gamma^{\mu} A_\mu (X) \Psi_f(X)] \, ; 
\label{eq:Lel}
\end{eqnarray}

ii) the Lagrangian of the fermion-axion interaction, 
when $k=A$, $g_A = C_f/(2f_a)$, $\Gamma_A j_A = \gamma^{\mu} \gamma_5 \partial_{\mu} a(X)$,  
$a(X)$ is the  quantized axion field, $f_a$ is the Peccei-Quinn symmetry violation 
scale, $C_f$ is the model dependent factor of order unity:
\begin{eqnarray}
{\cal L}(X) \, = \, \frac{C_f}{2 f_a} 
[\bar \Psi_f (X) \gamma^{\mu} \gamma_5 \Psi_f(X)] \partial_{\mu} a(X) \, , 
\label{eq:Laxion}
\end{eqnarray}

iii) the effective local Lagrangian of the four-fermion weak interaction, 
when $k=V$, $g_V = G_{\mathrm{F}} C_V/\sqrt{2}$ and $k=A$, 
$g_A = - G_{\mathrm{F}} C_A/\sqrt{2}$:  
\beq
{\cal L}(X) \, = \, \frac{G_{\mathrm{F}}}{\sqrt 2}\,
\big [ \bar \Psi_f (X) \gamma_{\alpha} (C_V - C_A \gamma_5) \Psi_f (X) \big ] \,
J_{\alpha}  (X)   \,, 
\label{eq:Llocal}
\eeq
where $J_{\alpha} (X) = \bar \nu (X) \gamma_{\alpha} (1- \gamma_5) \nu (X)$ 
is the current of left-handed neutrinos;
$C_V = \pm 1/2 + 2 \sin^2 \theta_W, \, C_A = \pm 1/2$, and 
$\theta_\mathrm{W}$ is the Weinberg angle. 
Here, the upper sign corresponds to neutrinos of the same flavor $f$
($\nu = \nu_f$), when there is an exchange reaction both of $W$ and $Z$ bosons. 
The lower sign corresponds to the case of another neutrino flavors ($\nu \ne \nu_f$), when there is only 
$Z$ boson exchange. 
The conditions of applicability of the effective Lagrangian~(\ref{eq:Llocal}) should be specified. 
First, it is the condition of relatively small momentum transfers, $|q^2| \ll m_W^2\,$, 
where $m_W$ is the $W$ boson mass. And second, the condition that additionally arises 
in an external magnetic field, is $e B \ll m_W^2\,$. We will consider physical situations where 
both of these conditions are satisfied.

%
\begin{figure}
\centerline{\includegraphics[width=8cm]{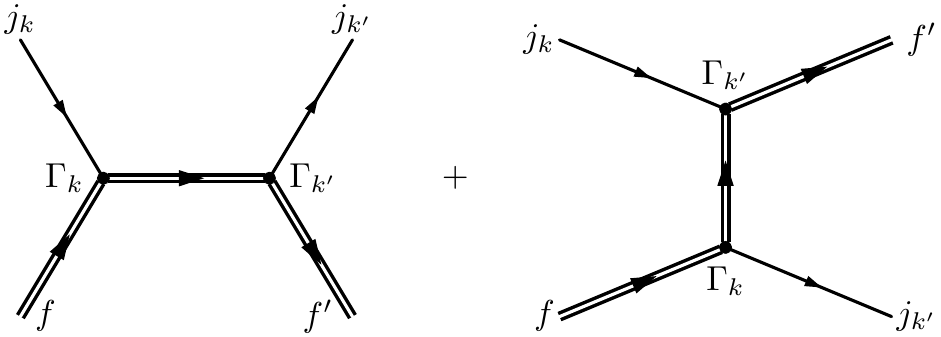}}
\caption{ The Feynman diagrams for the reaction $jf \to j^{\, \prime} f^{\, \prime}$. 
Double lines mean that the effects of an external field on the initial and final fermion states 
and on the fermion propagator are exactly taken into account.}
\label{fig:Diagjj}
\end{figure}
%

In a general case with the Lagrangian~(\ref{eq:L}), 
the $S$-matrix element in the tree approximation is described by the Feynman diagrams 
shown in Fig.~\ref{fig:Diagjj} and has the form
\beq                          
\label{eq:S1}
S^{s^{\, \prime} s}_{k^{\, \prime} k} &=& - g_k g_{k'}\int \dd^4 X \dd^4 Y 
\, \langle j_k (X) j_{k'} (Y) \rangle
\left [\bar \Psi^{s'}_{p',\ell'}(Y) \Gamma_{k'} 
\hat S(Y,X) 
\Gamma_k \Psi^{s}_{p,\ell}(X) \right ]
\\
\nonumber
&+&  (j_k, \Gamma_k \leftrightarrow j_k', \Gamma_k')\, .
\eeq

\noindent Here,  $p^{\mu} = (E_{\ell}, {\bf p})$ and $p^{\, \prime \mu} =
(E^{\, \prime}_{\ell'}, {\bf p}^{\, \prime})$ 
are the four-momenta of the initial and final fermions correspondingly, $X^{\mu} = (X_0, X_1, X_2, X_3)$, 
the currents between the angle brackets mean the matrix element between the corresponding initial 
and final states,
$\Psi^s_{p,\ell}(X)$ are the fermion wave functions in the presence of external magnetic field, 
where the subscript describes a state with definite components of the four-momentum $p$ 
and with the Landau level number $\ell$, while the superscript describes the spin state $s$. 

There exist several descriptions of the procedure of obtaining the fermion wave functions in the presence 
of an external magnetic field by solving the Dirac equation, see
e.g. Refs.~\refcite{Johnson:1949}--\refcite{Balantsev:2011} and 
also Refs.~\refcite{KM_Book_2003,KM_Book_2013}. 
In the most cases, the solutions are presented in the form with the upper two components 
of the bispinor corresponding to the fermion states with the spin projections 1/2 and -1/2
on the magnetic field direction.
Here, we have found it more convenient to use another representation of the fermion wave functions, 
being the eigenstates of the magnetic moment operator~\cite{Sokolov:1968,Melrose:1983a}. 
Some details on these wave functions are presented in~\ref{Append:A}. 

The currents $j_k$ in Eq.~(\ref{eq:S1}) can be expressed through
the amplitudes in the momentum space:
\beq
\label{eq:j_k}
j_k (X) = \frac{e^{-\ii(qX)}}{\sqrt{2q_0 V}} \, j_k(q) \,.
\eeq

\noindent We use the fermion propagator in the form of the sum over 
the Landau levels~\cite{Kuznetsov:2011,KM_Book_2013}:
\begin{equation}
S (X, X^{\,\prime}) = \sum\limits_{n=0}^{\infty} \; S_n (X, X^{\,\prime}) \,,
\label{eq:propagator_sum_n}
\end{equation}
\beq                          
&&S_n (X, X^{\,\prime}) = \frac{\ii}{2^n \, n!} \, \sqrt{\frac{\beta}{\pi}} \, 
\exp \left(- \, \beta \, \frac{X_1^2 + X_1^{\,\prime\,2}}{2} \right)
\nonumber\\[3mm]
&&\times 
\int\frac{\dd p_0 \, \dd p_y \, \dd p_z}{(2 \pi)^3 }
\frac{\eee^{- \, \ii \, \left( p \,(X - X^{\,\prime}) \right)_{\parallel}}}
{p_{\shortparallel}^2 - m_f^2 - 2 \, \beta \, n + \ii\,\varepsilon}\,
\nonumber\\[3mm]
&&\times 
\exp \left\{ - \, \frac{p_y^2}{\beta} 
- p_y \left[\,X_1 + X_1^{\,\prime} - \ii \, (X_2 - X_2^{\,\prime})\right] \right\} 
\nonumber\\[3mm]
&&\times 
\bigg\{ 
\left[ (p \gamma)_{\shortparallel} + m_f \right]
\left[ \varPi_- \, H_n (\xi) \, H_n (\xi^{\,\prime})  
+\varPi_+ \, 2 n \, H_{n-1} (\xi) \, H_{n-1} (\xi^{\,\prime}) 
\right]  
\nonumber\\[3mm] 
&& 
+ \, \ii \, 2 n \, \sqrt{\beta} \, \gamma^1 
\left[ \varPi_- \, H_{n-1} (\xi) \, H_n (\xi^{\,\prime}) 
-\varPi_+ \, H_n (\xi) \, H_{n-1} (\xi^{\,\prime}) 
\right]
\bigg\} ,
\label{eq:propn2}
\end{eqnarray}
where $\xi$ is defined by Eq.~(\ref{eq:xi}) and $\xi^{\,\prime}$ is obtained from $\xi$ by substituting 
$X_1 \to X_1^{\,\prime}$.

Hereafter we use the following notations:
four-vectors with the indices $\bot$ and $\parallel$
belong
to the Euclidean \{1, 2\} subspace and the Minkowski \{0, 3\} subspace correspondingly.
Then for arbitrary 4-vectors
$A_\mu$, $B_\mu$ one has
\beq
&&A_{\mprp}^\mu = (0, A_1, A_2, 0), \quad  A_{\mprl}^\mu = (A_0, 0, 0, A_3), \nonumber \\
&&(A B)_{\mprp} = (A \Lambda B) =  A_1 B_1 + A_2 B_2 , \nonumber \\
&&(A B)_{\mprl} = (A \widetilde \Lambda B) = A_0 B_0 - A_3 B_3, \nonumber
\end{eqnarray}

\noindent where the matrices
$\Lambda_{\mu \nu} = (\varphi \varphi)_{\mu \nu}$,\,
$\widetilde \Lambda_{\mu \nu} =
(\tilde \varphi \tilde \varphi)_{\mu \nu}$ are constructed with
the dimensionless tensor of the external
magnetic field, $\varphi_{\mu \nu} =  F_{\mu \nu} /B$,
and the dual tensor,
${\tilde \varphi}_{\mu \nu} = \frac{1}{2}
\varepsilon_{\mu \nu \rho \sigma} \varphi^{\rho \sigma}$.
The matrices $\Lambda_{\mu \nu}$ and  $\widetilde \Lambda_{\mu \nu}$
are connected by the relation
$\widetilde \Lambda^{\mu \nu} - \Lambda^{\mu \nu} =
g^{\mu \nu} = \rm{diag} (1, -1, -1, -1)$,
and play the roles of the metric tensors in the perpendicular ($\bot$)
and the parallel ($\parallel$) subspaces respectively.

After integration in Eq.~(\ref{eq:S1}) over $\dd^4X$ and $\dd^4Y$ 
we obtain
\beq
\label{eq:S2}                                  
&&S^{s^{\,\prime} s}_{k^{\, \prime} k} = \frac{\ii (2\pi)^3 
\delta^{(3)} (P - p^{\, \prime} - q^{\, \prime})}
{\sqrt{2q_0 V 2q^{\, \prime}_0 V 2 E_{\ell} L_y L_z 2 E^{\, \prime}_{\ell'}L_y L_z}}\, 
{\cal M}^{s^{\, \prime} s}_{k^{\, \prime} k} \, , 
\eeq
\noindent where    $\delta^3 (P - p^{\, \prime} - q^{\, \prime}) 
= \delta (P_0 - E^{\, \prime}_{\ell'}-q^{\, \prime}_0) 
\delta (P_y - p^{\, \prime}_y - q^{\, \prime}_y) 
\delta (P_z - p^{\, \prime}_z - q^{\, \prime}_z)$,  $P_\alpha = (p+q)_\alpha , 
\,\, \alpha =0,2,3$,  and the partial amplitudes 
${\cal M}^{s^{\, \prime} s}_{k^{\, \prime} k}$  can be  
presented in the following form:
\beq
&&{\cal M}^{s^{\, \prime} s}_{k^{\, \prime} k} = \frac{-
\exp{\left [-\ii\theta \right ]}}
{2\; \sqrt{M_\ell M_{\ell'} (M_\ell + m_f)(M_{\ell'} + m_f)}} 
\\
\nonumber
&&\times \bigg \{ \exp \left [\frac{\ii (q \varphi q^{\, \prime})}{2\beta} \right ]
\, \left [\frac{q_y + \ii q_x}{\sqrt{q^2_{\mprp}}} \right ]^{-\ell}
\, \left [\frac{q^{\, \prime}_y - \ii q^{\, \prime}_x}
{\sqrt{q^{\, \prime 2}_{\mprp}}} \right ]^{-\ell'} 
\\
\nonumber
&&\times \sum \limits_{n=0}^{\infty}
\; \left (\frac{(q\Lambda q^{\, \prime})-\ii (q\varphi q^{\, \prime})}
{\sqrt{q^2_{\mprp} q^{\, \prime 2}_{\mprp}}} \right )^n 
\frac{{\cal R}^{(1) s^{\, \prime} s}_{k^{\, \prime} k}}{P^2_{\mprl} - m_f^2 - 2\beta n}  
\\
\nonumber
&&+ (-1)^{\ell + \ell'} \exp \left [-\frac{\ii (q \varphi q^{\, \prime})}{2\beta} \right ]
\, \left [\frac{q^{\, \prime}_y + \ii q^{\, \prime}_x}{\sqrt{q^{\, \prime  2}_{\mprp}}} \right ]^{-\ell}
\, \left [\frac{q_y - \ii q_x}
{\sqrt{q^{2}_{\mprp}}} \right ]^{-\ell'} 
\\
\nonumber
&&\times \sum \limits_{n=0}^{\infty}
\; \left (\frac{(q\Lambda q^{\, \prime})+\ii (q\varphi q^{\, \prime})}
{\sqrt{q^2_{\mprp} q^{\, \prime 2}_{\mprp}}} \right )^n 
\frac{{\cal R}^{(2) s^{\, \prime} s}_{k k^{\, \prime}}}{P^{\, \prime 2}_{\mprl} - m_f^2 - 2\beta n} \bigg \}
\, ,
\label{eq:M11}
\eeq
\noindent where  $\theta = (q_x - q'_x)(p_y+p'_y)/(2\beta)$ is the general phase for both 
diagrams in Fig.~\ref{fig:Diagjj}, $P'_\alpha = (p-q')_\alpha$. 

The main part of the problem is to calculate the values
${\cal R}^{(1,2) s^{\, \prime}s}_{k^{\, \prime} k}$ which are expressed via 
the following Lorentz covariants in the $\{0, 3\}$-subspace
\beq
\label{eq:K1}
{\cal K}_{1\alpha} = \sqrt{\frac{2}{(p\widetilde \Lambda p^{\, \prime}) + 
M_\ell M_{\ell'}}} 
 \left \{M_\ell (\widetilde \Lambda p^{\, \prime})_\alpha + 
M_{\ell'} (\widetilde \Lambda p)_\alpha  \right \}\, ,
\eeq
\beq
\label{eq:K2}
{\cal K}_{2\alpha} = \sqrt{\frac{2}{(p\widetilde \Lambda p^{\, \prime}) + 
M_\ell M_{\ell'}}} 
 \left \{M_\ell (\widetilde \varphi p^{\, \prime})_\alpha + 
M_{\ell'} (\widetilde \varphi p)_\alpha  \right \}\, ,
\eeq
\beq
{\cal K}_{3} = \sqrt{2\left [(p\widetilde \Lambda p^{\, \prime}) + 
M_\ell M_{\ell'} \right]} \, ,  
\eeq
\beq
{\cal K}_4 = 
- \sqrt{\frac{2}{(p\widetilde \Lambda p^{\, \prime}) + M_\ell M_{\ell'}}}\, 
(p\widetilde \varphi p^{\, \prime}) \, .
\label{eq:K34}
\eeq

\noindent The following integrals appear in the calculations: 
\beq
\label{eq:Ilnnl}
&&\frac{1}{\sqrt{\pi}}\int \dd Z \, \eee^{-Z^2} 
H_n \!\! \left (Z + \frac{q_y + \ii q_x}{2\sqrt{\beta}} \right )   
 H_{\ell} \!\! \left (Z - \frac{q_y - \ii q_x}{2\sqrt{\beta}} \right ) 
\\[3mm]
\nonumber
&&= 2^{(n+\ell)/2} \sqrt{n ! \, \ell !} 
\left [\frac{q_y + \ii q_x}{\sqrt{q^2_{\mprp}}} \right ]^{n-\ell} 
\eee^{{q^2_{\mprp}}/{(4\beta)}} \,
{\cal I}_{n, \ell} \! \left (\frac{q^{2}_{\mprp}}{2 \beta} \right ) \, , 
\eeq
\noindent where, for $n \geqslant \ell$
\beq
\label{eq:Inl}
{\cal I}_{n, \ell} (x) = \sqrt{\frac{\ell !}{n !}} \; \eee^{-x/2} \, x^{(n-\ell)/2} L_\ell^{n-\ell} (x) \, ,
 \quad {\cal I}_{\ell, n} (x) = (-1)^{n-\ell} \, {\cal I}_{n, \ell} (x) \, ,
\eeq
\noindent and $L^k_n (x)$ are the generalized Laguerre polynomials~\cite{Gradshtein}.

Below, the results are presented for the values ${\cal R}^{(1,2) s^{\, \prime} s}_{k^{\, \prime} k}$ 
in the case when both vertices are of the pseudoscalar type, $k = k^{\, \prime} = P$. 
The total set of the values ${\cal R}^{(1,2) s^{\, \prime} s}_{k^{\, \prime} k}$ 
for the $jff$ interaction vertices of the scalar, pseudoscalar, vector and  
axial-vector types, in a constant uniform magnetic field of an arbitrary strength and in  
charged fermion plasma is presented in the extended paper~\cite{Kuznetsov:2013}. 

Hereafter we use the following definitions: 
${\cal I}_{n, \ell} \equiv {\cal I}_{n, \ell}
\left ({q^{2}_{\mprp}}/{(2 \beta)} \right )$ and 
${\cal I}^{\, \prime}_{n, \ell'} \equiv {\cal I}_{n, \ell'}
\left ({q^{\, \prime 2}_{\mprp}}/{(2 \beta)} \right )$. 
For definiteness, we further consider the fermion with a negative charge, $e_f = - |e_f|$. 

In the case when $j$ and $j^{\, \prime}$ are the pseudoscalar currents 
($k =  k^{\, \prime} = P$) we obtain
\beq
\nonumber
&&{\cal R}^{(1) ++}_{PP} = - g_P g_P^{\, \prime} j_P j_P^{\, \prime} \bigg \{ 2\beta 
\sqrt{\ell \ell^{\, \prime}} \left [({\cal K}_1 P) + m_f {\cal K}_3 \right ] 
{\cal I}^{\, \prime}_{n, \ell'} {\cal I}_{n, \ell} 
\\[3mm]
\label{eq:Rppupup}
&&+
(M_{\ell} + m_f) (M_{\ell'} + m_f) 
\left [({\cal K}_1 P) - m_f {\cal K}_3 \right ] 
 {\cal I}^{\, \prime}_{n-1, \ell'-1} {\cal I}_{n-1, \ell-1} 
\\[3mm]
\nonumber
&&- 2\beta \sqrt{n} {\cal K}_3 \big [\sqrt{\ell} (M_{\ell'} + m_f) 
{\cal I}^{\, \prime}_{n-1, \ell'-1} {\cal I}_{n, \ell} + 
\sqrt{\ell^{\, \prime}} (M_{\ell} + m_f) 
{\cal I}^{\, \prime}_{n, \ell'} {\cal I}_{n-1, \ell-1} \big ]  
\bigg \}\, ; 
\eeq
\beq
\nonumber
&&{\cal R}^{(1) +-}_{PP} = - \ii g_P g_P^{\, \prime} j_P j_P^{\, \prime} 
\bigg \{ \sqrt{2\beta \ell^{\, \prime}}\; (M_{\ell} + m_f) 
\left [({\cal K}_2 P) + m_f {\cal K}_4 \right ] 
{\cal I}^{\, \prime}_{n, \ell'} {\cal I}_{n, \ell} 
\\[3mm]
\label{eq:Rppupdown}
&&-
\sqrt{2\beta \ell}\; (M_{\ell'} + m_f) 
\left [({\cal K}_2 P) - m_f {\cal K}_4 \right ]
 {\cal I}^{\, \prime}_{n-1, \ell'-1} {\cal I}_{n-1, \ell-1} 
\\[3mm]
\nonumber
&&-\sqrt{2\beta n}\; {\cal K}_4 \big [ (M_{\ell} + m_f) (M_{\ell'} + m_f)
{\cal I}^{\, \prime}_{n-1, \ell'-1} {\cal I}_{n, \ell} - 
2\beta
\sqrt{\ell \ell^{\, \prime}}
{\cal I}^{\, \prime}_{n, \ell'} {\cal I}_{n-1, \ell-1} \big ]  
\bigg \}\, ; 
\eeq
\beq
\nonumber
&&{\cal R}^{(1) -+}_{PP} =   \ii g_P g_P^{\, \prime} j_P j_P^{\, \prime} 
\bigg \{ \sqrt{2\beta \ell}\; (M_{\ell'} + m_f) 
\left [({\cal K}_2 P) - m_f {\cal K}_4 \right ] 
{\cal I}^{\, \prime}_{n, \ell'} {\cal I}_{n, \ell} 
\\[3mm]
\label{eq:Rppdownup}
&&-
\sqrt{2\beta \ell^{\, \prime}}\; (M_{\ell} + m_f) 
\left [({\cal K}_2 P) + m_f {\cal K}_4 \right ]
{\cal I}^{\, \prime}_{n-1, \ell'-1} {\cal I}_{n-1, \ell-1} 
\\[3mm]
\nonumber
&&-
\sqrt{2\beta n}\; {\cal K}_4 \big [2\beta
\sqrt{\ell \ell^{\, \prime}} 
{\cal I}^{\, \prime}_{n-1, \ell'-1} {\cal I}_{n, \ell} - 
(M_{\ell} + m_f) (M_{\ell'} + m_f)
{\cal I}^{\, \prime}_{n, \ell'} {\cal I}_{n-1, \ell-1} \big ]  
\bigg \}\, ; 
\eeq
\beq
\nonumber
&&{\cal R}^{(1) --}_{PP} = - g_P g_P^{\, \prime} j_P j_P^{\, \prime} 
\bigg \{ (M_{\ell} + m_f) (M_{\ell'} + m_f)
\left [({\cal K}_1 P) - m_f {\cal K}_3 \right ]  
{\cal I}^{\, \prime}_{n, \ell'} {\cal I}_{n, \ell} 
\\[3mm]
\label{eq:Rppdowndown}
&&+
2\beta 
\sqrt{\ell \ell^{\, \prime}} \left [({\cal K}_1 P) + m_f {\cal K}_3 \right ]
 {\cal I}^{\, \prime}_{n-1, \ell'-1} {\cal I}_{n-1, \ell-1} 
\\[3mm]
\nonumber
&& -
2\beta \sqrt{n} {\cal K}_3 \big [\sqrt{\ell^{\, \prime}} (M_{\ell} + m_f) 
{\cal I}^{\, \prime}_{n-1, \ell'-1} {\cal I}_{n, \ell} + 
\sqrt{\ell} (M_{\ell'} + m_f) 
{\cal I}^{\, \prime}_{n, \ell'} {\cal I}_{n-1, \ell-1} \big ]  
\bigg \}\, . 
\eeq

To obtain the contributions ${\cal R}^{(2) s^{\, \prime} s}_{PP}$ from the second diagram of Fig.~\ref{fig:Diagjj}, 
the following replacements 
should be made in Eqs.~(\ref{eq:Rppupup})---(\ref{eq:Rppdowndown}): $P_{\alpha} \to P_{\alpha}^{\, \prime}$, 
${\cal I}_{m,n} \leftrightarrow {\cal I}_{m,n}^{\, \prime}$.

The results obtained for the case of the magnetic field of an arbitrary strength can 
be essentially simplified in several special cases.  
In~\ref{Append:B}, the set of expressions for the amplitudes in the limit of relatively strong field is presented, where  the 
initial and final fermions are on the ground Landau level,  
$\ell,\; \ell^{\, \prime} =0$, but the virtual electron can occupy an arbitrary Landau level, $n \ne 0$. 

One more case when the amplitudes can be essentially simplified is 
the process of a coherent scattering of the generalized current $j$ 
off the real fermions of magnetized plasma without change of their states (the ``forward'' scattering). 
The set of expressions for the amplitudes in this case is presented in~\ref{Append:C}. 


\section{Neutrino luminosity}
\label{Sec:4a} 


As an illustration of the results obtained, let us construct the amplitude of the neutrino-antineutrino pair 
photoproduction, $\gamma e \to e \nu \bar \nu$, 
in a strongly magnetized cold plasma when the temperature $T$ is the smallest 
parameter of the problem, i.e., $T \ll \mu_e -m_e$ ($\mu_e$ is the chemical potential of the electron gas, 
$m_e$ is the electron mass) with
taking account of a possible resonance on a virtual electron.

At the same time, our main goal is to obtain the expression for the neutrino emissivity 
caused by the process $\gamma e \to e \nu \bar \nu$. 
In turn, the neutrino emissivity can be defined as the zero component of the 
four-vector of the energy-momentum carried away by the neutrino pair due to
this process from a unit volume of plasma per unit time. 
Here, we neglect the inverse effect of the energy and momentum loss on the state of plasma. 
The neutrino emissivity can be represented in the form~\cite{Yakovlev2000}:
\beq
\label{eq:Q}
Q_{\gamma e \to e \nu \bar \nu} &=& \frac{1}{L_x}\; \sum\limits_{\ell, \ell' = 0}^{\infty} 
\int \frac{\dd^3 k}{(2\pi)^3 \, 2 q_0} \, f_\gamma (q_0) \, 
\frac{\dd^2 p }{(2\pi)^2 \, 2 E_{\ell}} \, 
f_e (E_{\ell}) \,  
\\[3mm]
\nonumber
&\times&
\frac{\dd^2 p'}{(2\pi)^2 \, 2 E^{\, \prime}_{\ell'} } \, \left [1-f_e (E^{\, \prime}_{\ell'}) \right ] \,  
\frac{\dd^3 p_1}{(2\pi)^3 \, 2 \varepsilon_1} \,
\frac{\dd^3 p_2}{(2\pi)^3 \, 2 \varepsilon_2} \,  q^{\, \prime}_0\,
\\[3mm]
\nonumber
&\times&
(2 \pi)^3  \, \delta^3 (P - p^{\, \prime} - q^{\, \prime}) 
|{\cal M}_{\gamma e \to e \nu \bar \nu}|^2 \, ,
\eeq
where $f_\gamma (q_0) =  \left [\eee^{q_0/T} - 1 \right ]^{-1}$ is the equilibrium 
distribution function of an initial photon with the four-vector $q^{\mu} = (q_0, {\bf k})$; 
 $f_e (E_{\ell})$ and $f_e (E^{\, \prime}_{\ell'})$ are the equilibrium distribution functions of
initial and final electrons, respectively, 
$f_e (E_{\ell}) = \left [\eee^{(E_{\ell} - \mu_e)/T} + 1 \right ]^{-1}$;  
$q^{\, \prime}_0 = \varepsilon_1 + \varepsilon_2$ is the neutrino pair energy, 
$\varepsilon_{1,2} = |{\bf p}_{1,2}|$; $\dd^2 p = \dd p_y \dd p_z$; 
$V = L_x L_y L_z$ is the plasma volume.

In calculating the amplitude ${\cal M}_{\gamma e \to e \nu \bar \nu}$ of the 
process $\gamma e \to e \nu \bar \nu$, 
we consider the case of relatively small momentum transfers 
compared with the $W$ boson mass, $|q^{\, \prime 2}| \ll m_W^2\,$. 
Then the corresponding interaction Lagrangian can be written as follows, see Eq.~(\ref{eq:Llocal}):
\beq
{\cal L}  =  \frac{G_{\mathrm{F}}}{\sqrt 2}\,
\big [ \bar \Psi \gamma_{\alpha} (C_V - C_A \gamma_5) \Psi \big ] \,
\big [\bar \nu \gamma_{\alpha} (1- \gamma_5) \nu \big ]  
+  e (\bar \Psi \gamma_\alpha \Psi) \, A_\alpha \, , 
\label{eq:Lgammanu}
\eeq
where $A_{\alpha}$ is the four-potential of the photon field.

Comparing~(\ref{eq:Lgammanu}) with the Lagrangian of the general form~(\ref{eq:L}) 
we find that the amplitude squared of the 
process $\gamma e \to e \nu \bar \nu$ can be represented as:
\beq
\label{eq:ampl1}
|{\cal M}_{\gamma e \to e \nu \bar \nu}|^2 = \sum\limits_{s',s} 
|{\cal M}_{VV}^{s's} + {\cal M}_{AV}^{s's}|^2 \, ,
\eeq
and in formulas~(\ref{eq:Mn00}), (\ref{eq:Rvvn01})--(\ref{eq:Rvan0}) 
one should put $m_f = m_e$,  
$g_V = e$, where $e>0$ is the elementary charge, 
$j_\alpha = \epsilon_\alpha$ is the initial photon polarization vector, 
$g^{\, \prime}_V = G_{\mathrm{F}} C_V/\sqrt{2}$, 
$g^{\, \prime}_A = - G_{\mathrm{F}} C_A/\sqrt{2}$, $j_\alpha^{\, \prime} 
= \bar \nu \gamma_{\alpha} (1- \gamma_5) \nu$.

It should be noted that the virtual electron resonance occurs only in the $s$ channel 
diagram (the first diagram in Fig.~\ref{fig:Diagjj}). 
Nevertheless, even with the simplification caused by the resonance behavior, the  
problem under consideration is still enough cumbersome, because the charged fermions can occupy 
arbitrary Landau levels. 
The problem could be significantly simplified in the physical conditions of magnetars.  
Indeed, in the outer crust of a magnetar, the following hierarchy of parameters should exist:~\cite{Yakovlev2000} 
$eB \gg m_e^2,\, \mu_e^2, \, T^2$. 
Thus, the electron plasma can be considered as a strongly magnetized one, and under 
these assumptions one can approximately assume that the initial and the final electrons 
would occupy the ground Landau level ($\ell = \ell' = 0$), while the virtual electron 
can occupy an arbitrary Landau level. 

In our case, $s'=s=-1$ and the amplitude squared~(\ref{eq:ampl1}) takes the form
\beq
\label{eq:ampl2}
&& |{\cal M}_{\gamma e \to e \nu \bar \nu}|^2  
= \left | \sum \limits_{n=1}^{\infty}
\; \left (\frac{(q\Lambda q^{\, \prime})-
\ii (q\varphi q^{\, \prime})}
{\sqrt{q^2_{\mprp} q^{\, \prime 2}_{\mprp}}} \right )^n 
 \frac{{\cal R}_{n}}{P^2_{\mprl} - m_e^2 - 2\beta n} \right |^2 \, , 
\eeq
where 
\beq
\label{eq:Rn}
{\cal R}_{n} = \frac{1}{n!} \, \left (\frac{q^2_{\mprp}}{2 \beta} \right )^{n/2} \, \left (\frac{q^{\, \prime 2}_{\mprp}}{2 \beta} \right )^{n/2} 
\,  \exp{\left [-\frac{q_{\mprp}^2+q^{\, \prime 2}_{\mprp}}
{4 \beta}\right ]} \, \left (R^{(1)}_{VV} +  R^{(1)}_{AV}\right ) \, , 
\eeq
and the functions $R^{(1)}_{VV}$ and $R^{(1)}_{AV}$ are defined by Eqs.~(\ref{eq:Rvvn01}) 
and~(\ref{eq:Ravn0}).

To accurately take into account the resonance behavior
in the process $\gamma e \to e \nu \bar \nu$, it is necessary to calculate 
radiative corrections to the electron mass, caused by the combined action of a 
magnetic field and plasma. This calculation is a separate challenge. 
However, because of the smallness of these corrections, we can approximately replace 
$m_e^2 \to m_e^2 - \ii P_0 \Gamma_n$ in the denominator of Eq.~(\ref{eq:ampl2}). 

As it was already noted, the main contribution to the amplitude arises from the resonance region, 
so that we can approximately replace the corresponding part of Eq.~(\ref{eq:ampl2}) by the
$\delta$ function:
\beq
\nonumber
&&|{\cal M}_{\gamma e \to e \nu \bar \nu}|^2 \simeq   \sum \limits_{n=1}^{\infty} \; 
 \frac{|{\cal R}_{n}|^2}{(P^2_{\mprl} - m_e^2 - 2\beta n)^2 + P_0^2 \Gamma_n^2}
\\
\label{eq:ampl3}
&& \simeq  \sum \limits_{n=1}^{\infty} \; 
\frac{\pi}{P_0 \Gamma_n}   \, 
\delta (P^2_{\mprl} - m_e^2 - 2 \beta n) \, |{\cal R}_{n}|^2 \, ,
\eeq
where $\Gamma_n$ is the total width of the change of the electron state. 
This width can be represented in the form~\cite{Weldon:1983} 
\beq
\label{eq:weldon}
\Gamma_n = \Gamma^{abs} + \Gamma^{cr} \simeq \Gamma^{cr}_{e_0 \gamma \to e_n} 
\left [1+ \eee^{(E^{\, \prime \prime}_n - \mu_e)/T} \right ]  \, .
\eeq
Here 
\beq
\nonumber
&&\Gamma^{cr}_{e_0 \gamma \to e_n}  = \frac{1}{2 E''_n} \, 
\int \frac{\dd^3 k }{2 q_0 (2\pi)^3} \, f_\gamma (q_0) \, 
\frac{\dd^2 p }{2 E_0 (2\pi)^2} \, f_e (E_0) \, 
\\
\label{eq:gammacr1}
&&\times (2\pi)^3 \, \delta^3 (P - p^{\, \prime \prime})
\, |{\cal M}_{e_0 \gamma  \to e_n}|^2  
\eeq 
is the width of the electron creation in the $n$th Landau level.

With taking account of Eq.~(\ref{eq:weldon}), the amplitude squared of the process 
$\gamma e \to e \nu \bar \nu$ takes the form:
\beq
\label{eq:resamp2}
&&|{\cal M}_{\gamma e \to e \nu \bar \nu}|^2 
=  \sum\limits_{n=1}^{\infty} \; 
\int \frac{\dd^2 p''}{(2 \pi)^2 \, 2 E''_n} \, 
(2 \pi)^3 \, 
\delta^3 (P - p^{\, \prime \prime}) \, \frac{|{\cal R}_n|^2}{2 E''_n \, \Gamma_n} 
\\
\nonumber
&&= \sum\limits_{n=1}^{\infty} \; 
\int \frac{\dd^2 p''}{(2 \pi)^2 \, 2 E''_n} \, f_e (E''_n)\, 
(2 \pi)^3 \, 
\delta^3 (P - p^{\, \prime \prime}) \, \frac{|{\cal R}_n|^2}{2 E''_n \, 
\Gamma^{cr}_{e_0 \gamma \to e_n}}
\, .
\eeq
Here we have used the property of the $\delta$ function:
\beq
\delta (P^2_{\mprl} - m_e^2 - 2 \beta n) = \frac{1}{2 E''_n} \, \delta (P_0 - E''_n) \, , 
\eeq
where $E''_n = \sqrt{p^{\, \prime \prime 2}_z + m_e^2 + 2 \beta n}$.

On the other hand, in the case of resonance the expression for $|R_n|^2$ being averaged over 
the photon polarizations can be factored in the strong field limit $\beta \gg m_e^2$
as follows (see, for example, Ref.~\refcite{Latal:86}):
\beq
\label{eq:factor}
|{\cal R}_n|^2 = |{\cal M}_{e_0 \gamma  \to e_n}|^2 \, |{\cal M}_{e_n \to e_0 \nu \bar \nu}|^2 \, ,
\eeq
where 
\beq
\nonumber
&&|{\cal M}_{e_0 \gamma  \to e_n}|^2 = \frac{8\pi \alpha}{n!} \, 
\exp{\left (-\frac{q^2_{\mprp}}{2\beta} \right )}
\left (\frac{q^2_{\mprp}}{2\beta} \right )^n \, M_n^2 \, (p \tilde \Lambda q)  \, 
\\
\label{eq:factor1}
&& \times \sum\limits_{\lambda = 1}^{3} 
\bigg [ \left |A_1^{(\lambda)} \right |^2 + 
\left |A_2^{(\lambda)} + \sigma A_3^{(\lambda)} \right |^2 \bigg ]
\eeq
is the amplitude of the absorption of a photon in the 
process $e_0 \gamma \to e_n$, when an electron passes from the ground Landau level 
to a higher Landau level $n$. 
The parameter $\sigma = (p \tilde \varphi q)/(p \tilde \Lambda q) = \pm 1$  
determines the direction of the photon propagation with respect to the magnetic field direction. 
$A_i^{(\lambda)}$ are the expansion coefficients of the photon polarization vector 
$\epsilon_{\mu}^{(\lambda)}$ over the basis of the 4-vectors:
\beq
\label{eq:basis}
 b_{\mu}^{(1)} = (\varphi q)_\mu, \qquad
 b_{\mu}^{(2)} = (\tilde \varphi q)_\mu, 
\qquad
 b_{\mu}^{(3)} = q^2 \, (\Lambda q)_\mu - q_\mu \, q^2_{\mbox{\tiny $\bot$}}, 
\qquad b_{\mu}^{(4)} = q_\mu \,. 
\eeq 
Given the gauge invariance, one has:
\beq
\epsilon_{\mu}^{(\lambda)} = \sum\limits_{i = 1}^{3} A_i^{(\lambda)} \, b_{\mu}^{(i)} \, .
\eeq
Finally, the amplitude squared of the electron transition from the $n$th Landau level to 
the ground level with the creation of the neutrino-antineutrino pair takes the form: 

\beq
\nonumber
&&|{\cal M}_{e_n \to e_0 \nu \bar \nu} |^2 =  
\frac{G_{\mathrm{F}}^2}{n!} \, 
\exp{\left (-\frac{q^{\, \prime \, 2}_{\mprp}}{2\beta} \right )}
\left (\frac{q^{\, \prime \, 2}_{\mprp}}{2\beta} \right )^n \, 
\frac{M_n^2}{(p^{\, \prime} \tilde \Lambda q^{\, \prime})} \, \bigg \{ 
\left |C_V (p^{\, \prime} \tilde \Lambda j^{\, \prime}) - 
C_A  (p^{\, \prime} \tilde \varphi j^{\, \prime})\right |^2 
\\
\nonumber
&&+ \frac{(j^{\, \prime} \tilde \Lambda j^{\, \prime \, *})}
{q^{\, \prime \, 2}_{\mprp}} 
\, \left [C_V (p^{\, \prime} \tilde \Lambda q^{\, \prime}) - 
C_A  (p^{\, \prime} \tilde \varphi q^{\, \prime})\right ]^2 -  
\frac{2}{q^{\, \prime \, 2}_{\mprp}} \, \left [C_V (p^{\, \prime} 
\tilde \Lambda q^{\, \prime}) - 
C_A  (p^{\, \prime} \tilde \varphi q^{\, \prime})\right ] 
\\
\label{eq:factor1_}
&&\times  \mathrm{Re} \left ( (q^{\, \prime} \tilde \Lambda j^{\, \prime}) \,  
\left [C_V (p^{\, \prime} \tilde \Lambda j^{\, \prime \, *}) - 
C_A  (p^{\, \prime} \tilde \varphi j^{\, \prime \, *}) \right ] 
\right ) \bigg \}.
\eeq

Substituting Eq.~(\ref{eq:resamp2}) into the expression for the luminosity~(\ref{eq:Q}), 
and taking into account Eqs.~(\ref{eq:gammacr1}) and~(\ref{eq:factor}), we obtain:
\beq
\label{eq:Qtot}
Q_{\gamma e_0 \to e_0 \nu \bar \nu} = \sum\limits_{n=1}^{\infty}  Q_{e_n \to e_0 \nu \bar \nu} \, ,
\eeq
where 
\beq
\nonumber
&&Q_{e_n \to e_0 \nu \bar \nu} =  \frac{1}{L_x}\; \int 
\frac{\dd^2 p^{\, \prime \prime} }{(2\pi)^2 \, 
2 E_n^{\, \prime \prime}} \, f_e (E_n^{\, \prime \prime}) \,  
\frac{\dd^2 p'}{(2\pi)^2 \, 2 E^{\, \prime} } \, 
\left [1-f_e (E^{\, \prime}) \right ] 
\\
\label{eq:Qnusynh}
&&\times  
\frac{\dd^3 p_1}{(2\pi)^3 \, 2 \varepsilon_1} \,
\frac{\dd^3 p_2}{(2\pi)^3 \, 2 \varepsilon_2} \,  
q^{\, \prime}_0\,
(2 \pi)^3  \, \delta^3 (p'' - p^{\, \prime} - q^{\, \prime}) 
|{\cal M}_{e_n \to e_0 \nu \bar \nu}|^2 \, ,
\eeq
is the neutrino luminosity due to the process $e_n \to e_0 \nu \bar \nu$. 
This result coincides, up to notation, with the result of Ref.~\refcite{Yakovlev2000}.

\section{Discussion}	
\label{Sec:Discussion}

In this paper, we have calculated the tree-level two-point amplitudes for the transitions 
$jf \to j^{\, \prime} f^{\, \prime}$ in a constant uniform magnetic field of an 
arbitrary strength, and in
charged fermion plasma, for generalized vertices of the scalar, pseudoscalar, vector and   
axial vector types.
It is remarkable, that all the amplitudes obtained are manifestly Lorentz invariant, due to the choice 
of the Dirac equation solutions as the eigenfunctions of the covariant operator $\hat{\mu}_z$. 
In this case, partial contributions to an amplitude from the channels with different 
fermion polarization states 
are calculated separately, by direct multiplication of the bispinors and the Dirac matrices. 
This approach is an alternative to the method where the amplitudes squared are calculated, 
with summation over the fermion polarization states, and with using the fermion density matrices, 
see, e.g. Refs.~\refcite{Andreev:2010,Gvozdev:2012}.
However, the use of the density matrix in a magnetic field, as is usually done 
in the absence of a magnetic field, in the case 
of the two-vertex processes leads to extreme difficulties in analytical calculations.

The set of the amplitudes for the transitions $jf \to j^{\, \prime} f^{\, \prime}$ 
in a constant uniform magnetic field 
of an arbitrary strength, and in charged fermion plasma, presented in this paper, 
can be used as a reference book in the investigations of the quantum processes 
in external active media.
The field effects are taken into account exactly, because 
exact solutions of the Dirac equation are used. Owing to this, the expression 
obtained here for the amplitude is quite general; in particular, 
it can be widely used to analyze various physical phenomena and processes 
in a magnetic field and in plasma.  
The amplitudes ${\cal M}_{SS}$ and ${\cal M}_{PP}$, 
which are diagonal in the generalized currents, differ only 
in factors from the 
external-medium-induced contributions to the mass operators 
of the corresponding scalar and pseudoscalar fields. 
The amplitude ${\cal M}_{VV}$ defines, for example, 
the medium-induced part of the photon polarization operator.
The amplitudes ${\cal M}_{VV}$ and ${\cal M}_{VA}$ 
describe the process amplitude for the radiative transition of a massless neutrino
$\nu \to \nu \gamma$. 
Similarly, one can obtain the amplitudes for the axion decay $a\to\nu\bar{\nu}$ 
and for axion--photon oscillations by means of the corresponding 
substitutions.

Furthermore, the results obtained can be used to analyze 
the reactions with a possible resonance on the virtual electron (see e.g. Ref.~\refcite{Rum:2013}). 
It is well known that the processes of this type play an important role in the magnetospheres 
of isolated neutron stars, providing the production of 
$e^+e^-$ plasma~\cite{Beloborodov:2007}.

Although the obtained formulas look quite cumbrous, there are certain areas of their application. We emphasize that these formulas are derived for a general case, namely, for arbitrary values of the magnetic field, therefore, they can recover the results within the limits of weak and superstrong fields. Further, in this general form the formulas definitely may be used 
for numerical calculations.

\section*{Acknowledgments}

We are grateful to M.\,V.~Chistyakov, A.\,A.~Gvozdev, and I.\,S.~Ognev for useful discussions.

The study was performed with the support by the Project No.~92 within the base part of the State Assignment 
for the Yaroslavl University Scientific Research, and was supported in part by the 
Russian Foundation for Basic Research (Project No. \mbox{14-02-00233-a}).


\appendix

\section{Solutions of the Dirac equation in an external magnetic field}
\label{Append:A}


In this Appendix, we present the fermion wave functions as the solutions of the Dirac equation in the presence 
of an external magnetic field, and simultaneously as the eigenfunctions of the magnetic moment operator~\cite{Sokolov:1968,Melrose:1983a}.

In Ref.~\refcite{Sokolov:1968}, an operator was introduced which was called the generalized spin 
tensor of the third rank. In modern standard notations, the operator takes the 
form\footnote{It should be noted that in Ref.~\refcite{Sokolov:1968}, the covariant bilinear forms 
were constructed of Dirac matrices by inserting them not between bispinors $\bar\psi$ and $\psi$ 
as accepted in modern literature,~\cite{Peskin:1995} but between bispinors $\psi^{\dagger}$ and $\psi$.}
\begin{eqnarray}
{\rm F}_{\mu \nu \lambda} = - \frac{\ii}{2} \left( P_\lambda \gamma_0 \sigma_{\mu \nu} 
+ \gamma_0 \sigma_{\mu \nu} P_\lambda \right),
\label{eq:Fgen}
\end{eqnarray}
where $\sigma_{\mu \nu} = (\gamma_\mu \gamma_\nu - \gamma_\nu \gamma_\mu)/2$, and 
$P^\lambda = \ii \partial^\lambda - e_f \, A^\lambda = \left( \ii \partial_0 - e_f \, A_0 \,, 
- \ii {\bs \nabla} - e_f {\bs A} \right)$ is the generalized four-momentum operator with $A^\lambda$ 
being the four-potential of an external magnetic field. 
Taking the component ${\rm F}_{\mu \nu 0}$ of the operator~(\ref{eq:Fgen}) and taking into 
account that in the Schr\"odinger form of the Dirac equation one has $\ii \partial_0 = H$, where 
$H = \gamma_0 \left( {\bs \gamma} {\bs P} \right) + m_f \, \gamma_0 + e_f A_0$ 
is the Dirac Hamiltonian, one can construct the vector operator
\begin{eqnarray}
{\mu}_i = - \frac{1}{2} \, \varepsilon_{ijk} \, {\rm F}_{jk0} \,, 
\label{eq:mu_i}
\end{eqnarray}
where $\varepsilon_{ijk}$ is the Levi-Civita symbol. This is the magnetic moment 
operator,~\cite{Sokolov:1968,Melrose:1983a} which can be presented in the form
\begin{eqnarray}
{\bs \mu} = m_f {\bs \Sigma} - \ii \gamma_0 \gamma_5 [{\bs \Sigma} \times \hat{\bs P}] \,. 
\label{eq:mu_vec}
\end{eqnarray}
It is straightforward to show that the components of the operator~(\ref{eq:mu_vec}) 
commute with the Hamiltonian, i.e. $H$ and ${\mu}_z$ have common eigenfunctions. 
In the non-relativistic limit, the operator~(\ref{eq:mu_vec}) is transformed to the ordinary Pauli magnetic moment
operator, thus having an obvious physical interpretation. 

It appears to be convenient to use the fermion wave functions as the 
eigenstates of the operator ${\mu}_z$~\cite{Sokolov:1968,Melrose:1983a} 
\begin{eqnarray}
{\mu}_z = m_f \Sigma_z - \ii \gamma_0 \gamma_5 [{\bs \Sigma} \times {\bs P}]_z \,, 
\label{eq:mu_z}
\end{eqnarray}
where ${\bs P} =  - \ii {\bs \nabla} - e_f {\bs A}$. 
We take the frame where the field is directed 
along the $z$ axis, and the Landau gauge where the four-potential is: $A^\lambda = (0, 0, x B, 0)$. 
 It is convenient to use the notation $\beta = |e_f| B$, and to introduce the sign of the fermion 
charge as $\eta = e_f/|e_f|$. 

Our choice of the Dirac equation solutions as the eigenfunctions of the operator $\hat{\mu}_z$ 
is caused by the following arguments. Calculations of the process widths with two or more vertices 
in an external magnetic field by the standard method, including the squaring the amplitude 
with all the Feynman diagrams and with summation or averaging over the fermion polarization states, 
contain significant computational difficulties. In this case, it is convenient to calculate partial 
contributions to the amplitude from the channels with different fermion polarization states and 
for each diagram separately, by direct multiplication of the bispinors and the Dirac matrices. 
The result, up to a total for both diagrams non-invariant phase, will have an explicit Lorentz 
invariant structure. On the contrary, the amplitudes obtained with using the solutions for 
a fixed direction of the spin, do not have Lorentz invariant structure. Only the amplitude squared, 
summed over the fermion polarization states, is manifestly Lorentz-invariant with respect to a boost 
along the magnetic field direction.

The fermion wave functions having the form
\beq
\label{eq:psie}
\Psi^s_{p,n}(X) = \frac{e^{-\ii(E_{n} X_0 - p_y X_2 - p_z X_3)}\; U^s_{n} (\xi)}
{\sqrt{4E_{n}M_n (E_{n} + M_n)(M_n + m_f) L_y L_z}} \, ,  
\eeq
where 
\beq
\label{eq:E_n,M_n}
E_n = \sqrt{M_n^2 + p_z^2}\, , \quad  M_n = \sqrt{m_f^2 + 2 \beta n}\, ,
\eeq
are the solutions of the equation
\beq
\label{eq:mu_z_Eq}
\hat{\mu}_z \,\Psi^s_{p,n}(X) = s \, M_n \, \Psi^s_{p,n}(X) \, , \quad s = \pm 1\,.
\eeq
It is convenient to present the bispinors $U^s_{n} (\xi)$ in the form of 
decomposition over the solutions for negative and positive fermion charge, $U^s_{n, \eta} (\xi)$:
\beq
\label{eq:U^s}
U^s_{n} (\xi) = \frac{1-\eta}{2} \, U^{s}_{n,-} (\xi) + \frac{1+\eta}{2} \, U^{s}_{n,+} (\xi) \,,
\eeq
where
\beq
\label{eq:U--}
&&U^{-}_{n,-} (\xi) = \left ( 
\begin{array}{c}
-\ii\sqrt{2\beta n} \, p_z V_{n-1} (\xi)\\[2mm]
(E_n + M_n)(M_n + m_f) V_n (\xi)\\[2mm]
-\ii\sqrt{2\beta n} (E_n + M_n) V_{n-1} (\xi)\\[2mm]
-p_z (M_n + m_f) V_n (\xi)
\end{array}
\right )  ,   
\\ [3mm]
\label{eq:U+-}
&&U^{+}_{n,-} (\xi) = \left ( 
\begin{array}{c}
(E_n + M_n) (M_n + m_f) V_{n-1} (\xi)\\[2mm]
-\ii\sqrt{2\beta n} \, p_z V_n (\xi)\\[2mm]
p_z (M_n + m_f) V_{n-1} (\xi)\\[2mm]
\ii \sqrt{2 \beta n} (E_n + M_n) V_n (\xi)
\end{array}
\right )\! , 
\eeq

\beq
\label{eq:U-+}
&&U^{-}_{n,+} (\xi) = \left ( 
\begin{array}{c}
\ii\sqrt{2\beta n} \, p_z V_{n} (\xi)\\[2mm]
(E_n + M_n)(M_n + m_f) V_{n-1} (\xi)\\[2mm]
\ii\sqrt{2\beta n} (E_n + M_n) V_{n} (\xi)\\[2mm]
-p_z (M_n + m_f) V_{n-1} (\xi)
\end{array}
\right ) \!\! ,   
\\ [3mm]
\label{eq:U++}
&&U^{+}_{n,+} (\xi) = \left ( 
\begin{array}{c}
(E_n + M_n) (M_n + m_f) V_{n} (\xi)\\[2mm]
\ii\sqrt{2\beta n} \, p_z V_{n-1} (\xi)\\[2mm]
p_z (M_n + m_f) V_{n} (\xi)\\[2mm]
-\ii \sqrt{2 \beta n} (E_n + M_n) V_{n-1} (\xi)
\end{array}
\right ) , 
\eeq
$V_n(\xi) \, (n = 0,1,2, \dots)$ 
are the normalized harmonic oscillator functions, which are expressed in terms of the 
Hermite polynomials $H_n(\xi)$ \cite{Gradshtein}:
\beq
\label{eq:V_n}
V_n (\xi) = \frac{\beta^{1/4}\eee^{-\xi^2/2}}{\sqrt{2^n n! \sqrt{\pi}}} \, H_n(\xi)\, ,
\eeq
\beq
\label{eq:xi}
\xi = \sqrt{\beta} \left (X_1 - \eta \frac{p_y}{\beta} \right ) .
\eeq
%

\section{The set of expressions for the amplitudes for ground Landau Level, $\ell = \ell' = 0$}
\label{Append:B}


In this Appendix, we consider the limit of relatively  strong field, where  the 
initial and final fermions are on the ground Landau level,  
$\ell,\; \ell^{\, \prime} =0$, but the virtual electron can occupy the arbitrary Landau level, 
$n \ne 0$. In this case 
$s=s'=-1$, 
$M_{\ell} = M_{\ell'}  = m_f$, and  

\beq
\label{eq:In0}
{\cal I}_{n, 0} (x) = \frac{1}{\sqrt{n !}} \; \eee^{-x/2} x^{n/2} \, , 
\quad
{\cal I}_{n-1, 0} (x) = \sqrt{\frac{n}{x}} \; {\cal I}_{n, 0} (x) \, .
\eeq
\noindent Denoting
${\cal R}^{(1,2) --}_{k^{\, \prime} k} \equiv (2m_f)^2 \, R^{(1,2)}_{k^{\, \prime} k}$ 
we obtain the following expressions for the amplitudes~(\ref{eq:M11}) with the 
vertices of the scalar, pseudoscalar, vector or  
axial vector types
\beq
\nonumber
&&{\cal M}^{--}_{k^{\, \prime} k} = - \exp{\left [-\ii\theta \right ]}
 \exp{\left [-\frac{q_{\mprp}^2+q^{\, \prime 2}_{\mprp}}
{4 \beta}\right ]} \; \sum \limits_{n=0}^{\infty} \frac{1}{n!} 
\\
\label{eq:Mn00}
&&\times  \bigg \{
  \exp \left [\frac{\ii (q \varphi q^{\, \prime})}{2\beta} \right ] 
\;  \left (\frac{(q\Lambda q^{\, \prime})-\ii (q\varphi q^{\, \prime})}
{2\beta} \right )^n  
\frac{R^{(1)}_{k^{\, \prime} k}}{P^2_{\mprl} - m_f^2 -2\beta n} 
\\
\nonumber
&&+ \exp \left [-\frac{\ii (q \varphi q^{\, \prime})}{2\beta} \right ]
   \left (\frac{(q\Lambda q^{\, \prime})+\ii (q\varphi q^{\, \prime})}
{2\beta} \right )^n  
\; \frac{R^{(2)}_{k k^{\, \prime}}}{P^{\, \prime \; 2}_{\mprl} 
- m_f^2 -2\beta n} \bigg \} \;  ,
\eeq
\noindent   where
\beq
\label{eq:Rssn10}
R^{(1)}_{SS} = g_S g_S^{\, \prime} j_S j_S^{\, \prime}  
\left [({\cal K}_1 P) + m_f {\cal K}_3 \right ] \, ;    
\eeq
\beq
\label{eq:Rssn20}
R^{(2)}_{SS} = R^{(1)}_{SS} (q \leftrightarrow -q^{\, \prime}) \, ; 
\eeq
%
\beq
\label{eq:Rpsn0}
R^{(1)}_{PS} =  g_S g_P^{\, \prime} j_S j_P^{\, \prime} 
\left [({\cal K}_2 P) - m_f {\cal K}_4 \right ] \, ;  
\eeq
\beq
\label{eq:Rspn0}
R^{(2)}_{SP} = - g_S g_P^{\, \prime} j_S j_P^{\, \prime} 
\left [({\cal K}_2 P^{\, \prime}) + m_f {\cal K}_4 \right ] \, ; 
\eeq
\beq
\nonumber
&&R^{(1)}_{VS} =  g_S g_V^{\, \prime} j_S 
\bigg \{ (P \tilde \Lambda j^{\, \prime}) {\cal K}_3 + 
(P \tilde \varphi j^{\, \prime})  {\cal K}_4 
+ m_f ({\cal K}_1 j^{\, \prime})  
\\
\label{eq:Rvsn0}
&&- \frac{2\beta n}{q_{\mprp}^{\, \prime 2}} \;  
\left [(q^{\, \prime} \Lambda j^{\, \prime}) - 
\ii (q^{\, \prime} \varphi j^{\, \prime}) \right ]  {\cal K}_3
 \bigg \} \, ; 
\eeq
\beq
\nonumber
&&R^{(2)}_{SV} =  g_S g_V^{\, \prime} j_S 
\bigg \{ (P^{\, \prime} \tilde \Lambda j^{\, \prime}) {\cal K}_3 - 
(P^{\, \prime} \tilde \varphi j^{\, \prime})  {\cal K}_4 
+ m_f ({\cal K}_1 j^{\, \prime})  
\\
\label{eq:Rsvn0}
&&+ 
\frac{2\beta n}{q_{\mprp}^{\, \prime 2}} \;  
\left [(q^{\, \prime} \Lambda j^{\, \prime}) + 
\ii (q^{\, \prime} \varphi j^{\, \prime}) \right ]  {\cal K}_3
 \bigg \} \, ; 
\eeq
%
\beq
\nonumber
&&R^{(1)}_{AS} =  g_S g_A^{\, \prime} j_S 
\bigg \{ (P \tilde \Lambda j^{\, \prime}) {\cal K}_4 + 
(P \tilde \varphi j^{\, \prime})  {\cal K}_3 
- m_f ({\cal K}_2 j^{\, \prime}) 
\\
\label{eq:Rasn0}
&&- \frac{2\beta n}{q_{\mprp}^{\, \prime 2}} \;  
\left [(q^{\, \prime} \Lambda j^{\, \prime}) - 
\ii (q^{\, \prime} \varphi j^{\, \prime}) \right ]  {\cal K}_4
 \bigg \} \, ; 
\eeq
\beq
\nonumber
&&R^{(2)}_{SA} = g_S g_A^{\, \prime} j_S 
\bigg \{  
(P^{\, \prime} \tilde \varphi j^{\, \prime})  {\cal K}_3 - 
(P^{\, \prime} \tilde \Lambda j^{\, \prime}) {\cal K}_4 
- m_f ({\cal K}_2 j^{\, \prime}) 
\\
\label{eq:Rsan0}
&&- \frac{2\beta n}{q_{\mprp}^{\, \prime 2}} \;  
\left [(q^{\, \prime} \Lambda j^{\, \prime}) + 
\ii (q^{\, \prime} \varphi j^{\, \prime}) \right ]  {\cal K}_4
 \bigg \}  \, ; 
\eeq
%
\beq
\label{eq:Rppn01}
R^{(1)}_{PP} = - g_P g_P^{\, \prime} j_P j_P^{\, \prime}  
\left [({\cal K}_1 P) - m_f {\cal K}_3 \right ] \, ; 
\eeq
\beq
\label{eq:Rppn02}
R^{(2)}_{PP} = R^{(1)}_{PP} (q \leftrightarrow -q^{\, \prime}) \, ; 
\eeq
%
\beq
\nonumber
&&R^{(1)}_{VP} = - g_P g_V^{\, \prime} j_P 
\bigg \{ (P \tilde \Lambda j^{\, \prime}) {\cal K}_4 + 
(P \tilde \varphi j^{\, \prime})  {\cal K}_3 
+ m_f ({\cal K}_2 j^{\, \prime}) 
\\
\label{eq:Rvpn0}
&&- \frac{2\beta n}{q_{\mprp}^{\, \prime 2}} \;  
\left [(q^{\, \prime} \Lambda j^{\, \prime}) - 
\ii (q^{\, \prime} \varphi j^{\, \prime}) \right ]  {\cal K}_4
 \bigg \} 
\eeq
\beq
\nonumber
&&R^{(2)}_{PV} =  g_P g_V^{\, \prime} j_P 
\bigg \{  
(P^{\, \prime} \tilde \varphi j^{\, \prime})  {\cal K}_3 - 
(P^{\, \prime} \tilde \Lambda j^{\, \prime}) {\cal K}_4 
+ m_f ({\cal K}_2 j^{\, \prime}) 
\\
\label{eq:Rpvn0}
&&- \frac{2\beta n}{q_{\mprp}^{\, \prime 2}} \;  
\left [(q^{\, \prime} \Lambda j^{\, \prime}) + 
\ii (q^{\, \prime} \varphi j^{\, \prime}) \right ]  {\cal K}_4
 \bigg \}  \, ; 
\eeq
%
\beq
\nonumber
&&R^{(1)}_{AP} =  - g_P g_A^{\, \prime} j_P 
\bigg \{ (P \tilde \Lambda j^{\, \prime}) {\cal K}_3 + 
(P \tilde \varphi j^{\, \prime})  {\cal K}_4 
- m_f ({\cal K}_1 j^{\, \prime})  
\\
\label{eq:Rapn0}
&& - \frac{2\beta n}{q_{\mprp}^{\, \prime 2}} \;  
\left [(q^{\, \prime} \Lambda j^{\, \prime}) - 
\ii (q^{\, \prime} \varphi j^{\, \prime}) \right ]  {\cal K}_3
 \bigg \} \, ; 
\eeq
\beq
\nonumber
&&R^{(2)}_{PA} =   g_P g_A^{\, \prime} j_P 
\bigg \{ (P^{\, \prime} \tilde \Lambda j^{\, \prime}) {\cal K}_3 - 
(P^{\, \prime} \tilde \varphi j^{\, \prime})  {\cal K}_4 
- m_f ({\cal K}_1 j^{\, \prime})  
\\
\label{eq:Rpan0}
&&+ \frac{2\beta n}{q_{\mprp}^{\, \prime 2}} \;  
\left [(q^{\, \prime} \Lambda j^{\, \prime}) + 
\ii (q^{\, \prime} \varphi j^{\, \prime}) \right ]  {\cal K}_3
 \bigg \} \, ; 
\eeq
%
\beq
\nonumber
&&R^{(1)}_{VV} = g_V g_V^{\, \prime} 
\bigg \{(P \tilde \Lambda j^{\, \prime}) ({\cal K}_1 j) +
(P \tilde \Lambda j) ({\cal K}_1 j^{\, \prime}) - 
(j \tilde \Lambda j^{\, \prime}) ({\cal K}_1 P)  
\\[3mm]
\nonumber
&& +m_f [(j \tilde \Lambda j^{\, \prime}){\cal K}_3 +  
(j \tilde \varphi j^{\, \prime}) {\cal K}_4 ]
\\[3mm]
\label{eq:Rvvn01}
&&  + 
\frac{2\beta n}{q_{\mprp}^2  q_{\mprp}^{\, \prime 2}} \; 
[(j \Lambda j^{\, \prime}) - \ii 
(j \varphi j^{\, \prime})]  
[({\cal K}_1 P) - m_f {\cal K}_3] 
[(q \Lambda q^{\, \prime}) + \ii (q \varphi q^{\, \prime})] 
\\[3mm]
\nonumber
&&-
\frac{2\beta n}{q_{\mprp}^{\, \prime 2}}\; ({\cal K}_1 j) \;   
[(q^{\, \prime} \Lambda j^{\, \prime}) - 
\ii (q^{\, \prime} \varphi j^{\, \prime})] -
\frac{2\beta n}{q_{\mprp}^{2}}\; ({\cal K}_1 j^{\, \prime}) \;   
[(q \Lambda j) + \ii (q \varphi j)] \bigg \} \, ; 
\eeq
%
\beq
\label{eq:Rvvn02}
R^{(2)}_{VV} = R^{(1)}_{VV} (q \leftrightarrow -q^{\, \prime}) \, ; 
\eeq
%
\beq
\nonumber
&&R^{(1)}_{AV} = g_V g_A^{\, \prime} 
\bigg \{(P \tilde \Lambda j^{\, \prime}) ({\cal K}_2 j) +
(P \tilde \Lambda j) ({\cal K}_2 j^{\, \prime}) - 
(j \tilde \Lambda j^{\, \prime}) ({\cal K}_2 P)  
\\[3mm]
\nonumber
&& -m_f [(j \tilde \Lambda j^{\, \prime}){\cal K}_4 +  
(j \tilde \varphi j^{\, \prime}) {\cal K}_3 ]
\\[3mm]
\label{eq:Ravn0}
&&  + 
\frac{2\beta n}{q_{\mprp}^2  q_{\mprp}^{\, \prime 2}} \; 
[(j \Lambda j^{\, \prime}) - \ii 
(j \varphi j^{\, \prime})]  
[({\cal K}_2 P) + m_f {\cal K}_4] 
[(q \Lambda q^{\, \prime}) + \ii (q \varphi q^{\, \prime})] 
\\[3mm]
\nonumber
&&-
\frac{2\beta n}{q_{\mprp}^{\, \prime 2}}\; ({\cal K}_2 j) \;   
[(q^{\, \prime} \Lambda j^{\, \prime}) - 
\ii (q^{\, \prime} \varphi j^{\, \prime})] -
\frac{2\beta n}{q_{\mprp}^{2}}\; ({\cal K}_2 j^{\, \prime}) \;   
[(q \Lambda j) + \ii (q \varphi j)] \bigg \} \, ; 
\eeq
\beq
\nonumber
&&R^{(2)}_{VA} = g_V g_A^{\, \prime} 
\bigg \{(P^{\, \prime} \tilde \Lambda j^{\, \prime}) ({\cal K}_2 j) +
(P^{\, \prime} \tilde \Lambda j) ({\cal K}_2 j^{\, \prime}) - 
(j \tilde \Lambda j^{\, \prime}) ({\cal K}_2 P^{\, \prime})  
\\[3mm]
\nonumber
&&+m_f [(j \tilde \Lambda j^{\, \prime}){\cal K}_4 -  
(j \tilde \varphi j^{\, \prime}) {\cal K}_3 ]
\\[3mm]
\label{eq:Rvan0}
&&  + 
\frac{2\beta n}{q_{\mprp}^2  q_{\mprp}^{\, \prime 2}} \; 
[(j \Lambda j^{\, \prime}) + \ii 
(j \varphi j^{\, \prime})]  
[({\cal K}_2 P^{\, \prime}) - m_f {\cal K}_4] 
[(q \Lambda q^{\, \prime}) - \ii (q \varphi q^{\, \prime})] 
\\[3mm]
\nonumber
&&+
\frac{2\beta n}{q_{\mprp}^{\, \prime 2}}\; ({\cal K}_2 j) \;   
[(q^{\, \prime} \Lambda j^{\, \prime}) + 
\ii (q^{\, \prime} \varphi j^{\, \prime})] +
\frac{2\beta n}{q_{\mprp}^{2}}\; ({\cal K}_2 j^{\, \prime}) \;   
[(q \Lambda j) - \ii (q \varphi j)] \bigg \}  \, ; 
\eeq
%
\beq
\nonumber
&&R^{(1)}_{AA} =  g_A g_A^{\, \prime} 
\bigg \{(P \tilde \Lambda j^{\, \prime}) ({\cal K}_1 j) +
(P \tilde \Lambda j) ({\cal K}_1 j^{\, \prime}) - 
(j \tilde \Lambda j^{\, \prime}) ({\cal K}_1 P)  
\\[3mm]
\nonumber
&&-m_f [(j \tilde \Lambda j^{\, \prime}){\cal K}_3 +  
(j \tilde \varphi j^{\, \prime}) {\cal K}_4 ]
\\[3mm]
\label{eq:Raan01}
&&  + 
\frac{2\beta n}{q_{\mprp}^2  q_{\mprp}^{\, \prime 2}} \; 
[(j \Lambda j^{\, \prime}) - \ii 
(j \varphi j^{\, \prime})]  
[({\cal K}_1 P) + m_f {\cal K}_3] 
[(q \Lambda q^{\, \prime}) + \ii (q \varphi q^{\, \prime})] 
\\[3mm]
\nonumber
&&-
\frac{2\beta n}{q_{\mprp}^{\, \prime 2}}\; ({\cal K}_1 j) \;   
[(q^{\, \prime} \Lambda j^{\, \prime}) - 
\ii (q^{\, \prime} \varphi j^{\, \prime})] -
\frac{2\beta n}{q_{\mprp}^{2}}\; ({\cal K}_1 j^{\, \prime}) \;   
[(q \Lambda j) + \ii (q \varphi j)] \bigg \} \; .
\eeq
%
\beq
\label{eq:Raan02}
R^{(2)}_{AA} = R^{(1)}_{AA} (q \leftrightarrow -q^{\, \prime}) \, ; 
\eeq

We note that the obtained results allow us to extract the limiting case $n=0$.
In particular, the amplitude ${\cal M}^{--}_{VV}$ containing the vector vertices only,
coinsides after corresponding transformations with the amplitude of the Compton process 
in a strong magnetic field, calculated earlier in Ref.~\refcite{RCh09} (see also Ref.~\refcite{RCh08} where 
the amplitude of the type ${\cal M}^{--}_{AV}$ was considered for the case $\ell' = \ell = n = 0$).
In addition, it is easy to check that the resulting amplitudes containing the vector vertices, are manifestly gauge invariant.
 

\section{The set of expressions for the amplitudes for forward scattering}
\label{Append:C}


For generalization of the results obtained in Ref.~\refcite{Borovkov:1999} to the case 
of magnetized plasma we consider the process of a coherent scattering of the generalized current $j$ 
off the real fermions without change of their states (the ``forward'' scattering). 
We remind that in this case we mean under the generalized current $j$ in the initial state only the field operator 
of a single particle, while the generalized current $j^{\, \prime}$ in the final state could be both 
the field operator of a single particle, and e.g. the neutrino current.  
In this case:  
$\ell = \ell^{\, \prime}$, $s=s^{\, \prime}$, $q^\mu = q^{\, \prime \mu}$, 
$p^\mu = p^{\, \prime \mu}$, ${\cal K}_{1 \alpha} = 2 (p \tilde \Lambda)_{\alpha}$,   
${\cal K}_{2\alpha} = 2 (\tilde \varphi p)_\alpha$, ${\cal K}_3 = 2 M_{\ell}$, ${\cal K}_4 = 0$.  

Since this is a coherent process, the total scattering amplitude is obtained by summing 
over all states of the medium fermions. 
We obtain the follwing results for the summed generalized amplitudes: 
\beq
\label{eq:FS1}
&&{\cal M}_{k^{\, \prime} k} = - \frac{\beta}{2 \pi^2} \; \sum_{\ell,n = 0}^{\infty} 
\int \frac{\dd p_z}{E_{\ell}} \; 
f_{f}(E_{\ell})\;   
\\
\nonumber
&&\times \left \{ \frac{{\cal D}^{(1)}_{k^{\, \prime} k}}
{(p+q)^2_{\mprl} - m_f^2 - 2\beta n} +  \frac{{\cal D}^{(2)}_{kk^{\, \prime}}}
{(p-q)^2_{\mprl} - m_f^2 - 2\beta n}\right \} \, ,
\eeq
\noindent where $f_{f}(E_{\ell}) = [1+\exp{(E_{\ell} - \mu_f)/T}]^{-1}$ is the fermion 
distribution function,  $T$ and $\mu_f$ are the temperature and the chemical potential of plasma 
correspondingly, 
%
\beq
\label{eq:FSS}
&&{\cal D}^{(1)}_{SS} = g_S g_S^{\, \prime} j_S j_S^{\, \prime} \left \{ [(q \widetilde \Lambda p) + 
2 \beta \ell + 2 m_f^2]
\right.
\\[2mm]
\nonumber
&&\times \left.
({\cal I}_{n,\ell}^2 + {\cal I}_{n-1,\ell-1}^2)  - 
 4 \beta \sqrt{n \ell}\, {\cal I}_{n,\ell} {\cal I}_{n-1,\ell-1} \right \} ;
\\[2mm]
\nonumber
&&{\cal D}^{(2)}_{SS} = {\cal D}^{(1)}_{SS} (q\to -q) \, ;
\eeq
%
%
\beq
\label{eq:FSP}
{\cal D}^{(1)}_{SP} = {\cal D}^{(2)}_{PS} = 
 g_S g_P^{\, \prime} j_S j_P^{\, \prime} (q \widetilde \varphi p)  
\left [{\cal I}_{n,\ell}^2 - {\cal I}_{n-1,\ell-1}^2 \right ] ;
\eeq
%
%
\beq
\label{eq:FSV}
&&{\cal D}^{(1)}_{VS} =  g_S g_V^{\, \prime} j_S m_f   
\left \{ [2 (p \widetilde \Lambda j^{\, \prime}) + (q \widetilde \Lambda j^{\, \prime})] 
\left [{\cal I}_{n,\ell}^2 + {\cal I}_{n-1,\ell-1}^2 \right ]  \right.
\\[3mm]
\nonumber
&&\left. - \sqrt{\frac{2 \beta \ell}{q_{\mprp}^2}} \, \left [[(q \Lambda j^{\, \prime}) + 
\ii (q \varphi j^{\, \prime})] {\cal I}_{n,\ell} {\cal I}_{n,\ell-1}  
+[(q \Lambda j^{\, \prime}) - 
\ii (q \varphi j^{\, \prime})] {\cal I}_{n-1,\ell} {\cal I}_{n-1,\ell-1}  \right]   \right. 
\\[3mm]
\nonumber
&&\left. - 
 \sqrt{\frac{2 \beta n}{q_{\mprp}^2}} \left  [[(q \Lambda j^{\, \prime}) +
\ii (q \varphi j^{\, \prime})] {\cal I}_{n,\ell-1} {\cal I}_{n-1,\ell-1} 
+ [(q \Lambda j^{\, \prime}) -
\ii (q \varphi j^{\, \prime})] {\cal I}_{n,\ell} {\cal I}_{n-1,\ell}  \right] 
\right \};
\eeq

%
\beq
\label{eq:FSV2}
&&{\cal D}^{(2)}_{SV} =  g_S g_V^{\, \prime} j_S m_f   
\left \{ [2 (p \widetilde \Lambda j^{\, \prime}) - (q \widetilde \Lambda j^{\, \prime})] 
\left [{\cal I}_{n,\ell}^2 + {\cal I}_{n-1,\ell-1}^2 \right ]  \right.
\\[3mm]
\nonumber
&&\left. + \sqrt{\frac{2 \beta \ell}{q_{\mprp}^2}} \, \left [[(q \Lambda j^{\, \prime}) - 
\ii (q \varphi j^{\, \prime})] {\cal I}_{n,\ell} {\cal I}_{n,\ell-1}  
+[(q \Lambda j^{\, \prime}) + 
\ii (q \varphi j^{\, \prime})] {\cal I}_{n-1,\ell} {\cal I}_{n-1,\ell-1}  \right]   \right. 
\\[3mm]
\nonumber
&&\left. + 
 \sqrt{\frac{2 \beta n}{q_{\mprp}^2}} \left  [[(q \Lambda j^{\, \prime}) -
\ii (q \varphi j^{\, \prime})] {\cal I}_{n,\ell-1} {\cal I}_{n-1,\ell-1} 
+ [(q \Lambda j^{\, \prime}) +
\ii (q \varphi j^{\, \prime})] {\cal I}_{n,\ell} {\cal I}_{n-1,\ell}  \right] 
\right \};
\eeq

%
%
\beq
\label{eq:FSA}
&&{\cal D}^{(1)}_{AS} =  g_S g_A^{\, \prime} j_S m_f 
[2 (p \widetilde \varphi j^{\, \prime}) + (q \widetilde \varphi j^{\, \prime})] 
\left [{\cal I}_{n,\ell}^2 - {\cal I}_{n-1,\ell-1}^2 \right ] ;
\\[2mm]
\nonumber
&&{\cal D}^{(2)}_{SA} = {\cal D}^{(1)}_{AS} (q\to -q) \, ;
\eeq

\beq
\nonumber
&&{\cal D}^{(1)}_{PP} = - g_P g_P^{\, \prime} j_P j_P^{\, \prime} 
\left \{ [(q \widetilde \Lambda p) + 
2 \beta \ell ] 
\left [{\cal I}_{n,\ell}^2 + {\cal I}_{n-1,\ell-1}^2 \right ]   
\right.
\\[2mm]
\label{eq:FPP}
&&\left. - 4 \beta \sqrt{n \ell}\, {\cal I}_{n,\ell} {\cal I}_{n-1,\ell-1} \right \} ;
\\[2mm]
\nonumber
&&{\cal D}^{(2)}_{PP} = {\cal D}^{(1)}_{PP} (q\to -q) \, ;
\eeq
\beq
\label{eq:FPV}
{\cal D}^{(1)}_{VP} = {\cal D}^{(2)}_{PV} 
= - g_P g_V^{\, \prime} j_P m_f (q \widetilde \varphi j^{\, \prime})  
\left [{\cal I}_{n,\ell}^2 - {\cal I}_{n-1,\ell-1}^2 \right ] ;
\eeq

\beq
\label{eq:FPA1}
&&{\cal D}^{(1)}_{AP} = - g_P g_A^{\, \prime} j_P m_f   
\left \{ (q \widetilde \Lambda j^{\, \prime}) 
\left [{\cal I}_{n,\ell}^2 + {\cal I}_{n-1,\ell-1}^2 \right ]   \right.
\\[3mm]
\nonumber
&&\left. +\sqrt{\frac{2 \beta \ell}{q_{\mprp}^2}} \, \left [[(q \Lambda j^{\, \prime}) + 
\ii (q \varphi j^{\, \prime})] {\cal I}_{n,\ell} {\cal I}_{n,\ell-1}  
+[(q \Lambda j^{\, \prime}) - 
\ii (q \varphi j^{\, \prime})] {\cal I}_{n-1,\ell} {\cal I}_{n-1,\ell-1}  \right]  \right. 
\\[3mm]
\nonumber
&&\left. - \sqrt{\frac{2 \beta n}{q_{\mprp}^2}} \, \left [[(q \Lambda j^{\, \prime}) +
\ii (q \varphi j^{\, \prime})] {\cal I}_{n,\ell-1} {\cal I}_{n-1,\ell-1}  
+ [(q \Lambda j^{\, \prime}) -
\ii (q \varphi j^{\, \prime})] {\cal I}_{n,\ell} {\cal I}_{n-1,\ell}  \right] 
\right \};
\eeq

\beq
\label{eq:FPA2}
&&{\cal D}^{(2)}_{PA} = - g_P g_A^{\, \prime} j_P m_f   
\left \{  (q \widetilde \Lambda j^{\, \prime}) 
\left [{\cal I}_{n,\ell}^2 + {\cal I}_{n-1,\ell-1}^2 \right ]   \right.
\\[3mm]
\nonumber
&&\left. -\sqrt{\frac{2 \beta \ell}{q_{\mprp}^2}} \, \left [[(q \Lambda j^{\, \prime}) - 
\ii (q \varphi j^{\, \prime})] {\cal I}_{n,\ell} {\cal I}_{n,\ell-1}  
- [(q \Lambda j^{\, \prime}) + 
\ii (q \varphi j^{\, \prime})] {\cal I}_{n-1,\ell} {\cal I}_{n-1,\ell-1}  \right]  \right. 
\\[3mm]
\nonumber
&&\left. - \sqrt{\frac{2 \beta n}{q_{\mprp}^2}} \, \left [[(q \Lambda j^{\, \prime}) -
\ii (q \varphi j^{\, \prime})] {\cal I}_{n,\ell-1} {\cal I}_{n-1,\ell-1}  
+ [(q \Lambda j^{\, \prime}) +
\ii (q \varphi j^{\, \prime})] {\cal I}_{n,\ell} {\cal I}_{n-1,\ell}  \right] 
\right \};
\eeq

\beq
\label{eq:FVV}
&&{\cal D}^{(1)}_{VV} =  g_V g_V^{\, \prime}    
\big \{ \left [ (p \widetilde \Lambda j) (P \widetilde \Lambda j^{\, \prime}) + 
(P \widetilde \Lambda j) (p \widetilde \Lambda j^{\, \prime})  - 
(j \widetilde \Lambda j^{\, \prime})[2 \beta \ell + (p \widetilde \Lambda q)] \right ] 
\\[3mm]
\nonumber
&& \times 
\left [{\cal I}_{n,\ell}^2 + {\cal I}_{n-1,\ell-1}^2 \right ]   
 + 4 \beta \sqrt{n\ell} \, (j \widetilde \Lambda j^{\, \prime}) 
{\cal I}_{n,\ell} {\cal I}_{n-1,\ell-1} 
\\[3mm]
\nonumber
&&- \sqrt{\frac{2 \beta \ell}{q_{\mprp}^2}} \, \left [
(P \widetilde \Lambda j) [(q \Lambda j^{\, \prime}) + 
\ii (q \varphi j^{\, \prime})]  + (P \widetilde \Lambda j^{\, \prime}) [(q \Lambda j) - 
\ii (q \varphi j)] \right] {\cal I}_{n,\ell} {\cal I}_{n,\ell-1}  
\\[3mm]
\nonumber
&&  - 
 \sqrt{\frac{2 \beta \ell}{q_{\mprp}^2}} \, \left [
(P \widetilde \Lambda j) [(q \Lambda j^{\, \prime}) - 
\ii (q \varphi j^{\, \prime})]  + 
 (P \widetilde \Lambda j^{\, \prime}) [(q \Lambda j) + 
\ii (q \varphi j)] \right] {\cal I}_{n-1,\ell-1} {\cal I}_{n-1,\ell}  
\\[3mm]
\nonumber
&& - 
\sqrt{\frac{2 \beta n}{q_{\mprp}^2}} \, \left [
(p \widetilde \Lambda j) [(q \Lambda j^{\, \prime}) - 
\ii (q \varphi j^{\, \prime})]  +  (p \widetilde \Lambda j^{\, \prime}) 
[(q \Lambda j) + 
\ii (q \varphi j)] \right] {\cal I}_{n,\ell} {\cal I}_{n-1,\ell}  
\\[3mm]
\nonumber
&& - \sqrt{\frac{2 \beta n}{q_{\mprp}^2}}  \left [
(p \widetilde \Lambda j) [(q \Lambda j^{\, \prime}) + 
\ii (q \varphi j^{\, \prime})]  +  (p \widetilde \Lambda j^{\, \prime}) 
[(q \Lambda j) - 
\ii (q \varphi j)] \right] {\cal I}_{n-1,\ell-1} {\cal I}_{n,\ell-1}  
\\[3mm]
\nonumber
&& + 
 [2 \beta \ell + (p \widetilde \Lambda q)] 
 \left [[(j \Lambda j^{\, \prime}) + 
\ii (j \varphi j^{\, \prime})] {\cal I}_{n,\ell-1}^2 + [(j \Lambda j^{\, \prime}) - 
\ii (j \varphi j^{\, \prime})] {\cal I}_{n-1,\ell}^2 \right ]  
\\[3mm]
\nonumber
&&+ 
 \frac{4 \beta \sqrt{n\ell}}{q_{\mprp}^2} \, 
[(q \Lambda j) (q \Lambda j^{\, \prime}) - 
(q \varphi j) (q \varphi j^{\, \prime})] {\cal I}_{n,\ell-1} {\cal I}_{n-1,\ell}
\big \};
\eeq

\beq
\nonumber
&&{\cal D}^{(2)}_{VV} = {\cal D}^{(1)}_{VV} (q\to -q ,\, \, j \leftrightarrow j^{\, \prime}) \, ;
\eeq

%
%
\beq
\nonumber
&&{\cal D}^{(1)}_{AV} =  g_V g_A^{\, \prime}    
\left \{ [(P \widetilde \Lambda j)  (j^{\, \prime} \widetilde \varphi p) + 
(P \widetilde \Lambda j^{\, \prime})  (j \widetilde \varphi p) 
\right.
\\[3mm]
\label{eq:FVA}
&&\left.
- 
(j \widetilde \Lambda j^{\, \prime}) (q \widetilde \varphi p) - 
m_f^2 (j \widetilde \varphi j^{\, \prime})] 
\left [{\cal I}_{n,\ell}^2 - {\cal I}_{n-1,\ell-1}^2 \right ]  \right.
\\[3mm]
\nonumber
&&\left. + \sqrt{\frac{2 \beta \ell}{q_{\mprp}^2}} \, \left [
(P \widetilde \varphi j) [(q \Lambda j^{\, \prime}) + 
\ii (q \varphi j^{\, \prime})]  + (P \widetilde \varphi j^{\, \prime}) 
[(q \Lambda j) - 
\ii (q \varphi j)] \right] {\cal I}_{n,\ell} {\cal I}_{n,\ell-1}  \right. 
\\[3mm]
\nonumber
&&\left. - \sqrt{\frac{2 \beta \ell}{q_{\mprp}^2}} \, \left [
(P \widetilde \varphi j) [(q \Lambda j^{\, \prime}) - 
\ii (q \varphi j^{\, \prime})]  + (P \widetilde \varphi j^{\, \prime}) 
[(q \Lambda j) + 
\ii (q \varphi j)] \right] {\cal I}_{n-1,\ell-1} {\cal I}_{n-1,\ell}  \right. 
\\[3mm]
\nonumber
&&\left. + \sqrt{\frac{2 \beta n}{q_{\mprp}^2}} \, \left [
(p \widetilde \varphi j) [(q \Lambda j^{\, \prime}) - 
\ii (q \varphi j^{\, \prime})]  +  (p \widetilde \varphi j^{\, \prime}) 
[(q \Lambda j) + 
\ii (q \varphi j)] \right] {\cal I}_{n,\ell} {\cal I}_{n-1,\ell}  \right. 
\\[3mm]
\nonumber
&&\left. - \sqrt{\frac{2 \beta n}{q_{\mprp}^2}} \, \left [
(p \widetilde \varphi j) [(q \Lambda j^{\, \prime}) + 
\ii (q \varphi j^{\, \prime})]  +  (p \widetilde \varphi j^{\, \prime}) 
[(q \Lambda j) - 
\ii (q \varphi j)] \right] {\cal I}_{n-1,\ell-1} {\cal I}_{n,\ell-1}  \right. 
\\[3mm]
\nonumber
&&\left. + (p \widetilde \varphi q) \left [[(j \Lambda j^{\, \prime}) + 
\ii (j \varphi j^{\, \prime})]{\cal I}_{n,\ell-1}^2 - [(j \Lambda j^{\, \prime}) - 
\ii (j \varphi j^{\, \prime})]{\cal I}_{n-1,\ell}^2 \right ] 
\right \};
\eeq
\beq
\nonumber
&&{\cal D}^{(2)}_{VA} =  g_V g_A^{\, \prime}    
\big \{ [(P^{\, \prime} \widetilde \Lambda j)  (j^{\, \prime} \widetilde \varphi p) + 
(P^{\, \prime} \widetilde \Lambda j^{\, \prime})  (j \widetilde \varphi p) 
\\[3mm]
\label{eq:FVA2}
&&+ 
(j \widetilde \Lambda j^{\, \prime}) (q \widetilde \varphi p) - 
m_f^2 (j \widetilde \varphi j^{\, \prime})] 
\left [{\cal I}_{n,\ell}^2 - {\cal I}_{n-1,\ell-1}^2 \right ]  
\\[3mm]
\nonumber
&& - \sqrt{\frac{2 \beta \ell}{q_{\mprp}^2}} \, \left [
(P^{\, \prime} \widetilde \varphi j^{\, \prime}) [(q \Lambda j) + 
\ii (q \varphi j)]  + (P^{\, \prime} \widetilde \varphi j) 
[(q \Lambda j^{\, \prime}) - 
\ii (q \varphi j^{\, \prime})] \right] {\cal I}_{n,\ell} {\cal I}_{n,\ell-1}   
\\[3mm]
\nonumber
&& + \sqrt{\frac{2 \beta \ell}{q_{\mprp}^2}} \, \left [
(P^{\, \prime} \widetilde \varphi j^{\, \prime}) [(q \Lambda j) - 
\ii (q \varphi j)]  + (P^{\, \prime} \widetilde \varphi j) 
[(q \Lambda j^{\, \prime}) + 
\ii (q \varphi j^{\, \prime})] \right] {\cal I}_{n-1,\ell-1} {\cal I}_{n-1,\ell}   
\\[3mm]
\nonumber
&& - \sqrt{\frac{2 \beta n}{q_{\mprp}^2}} \, \left [
(p \widetilde \varphi j^{\, \prime}) [(q \Lambda j) - 
\ii (q \varphi j)]  +  (p \widetilde \varphi j) 
[(q \Lambda j^{\, \prime}) + 
\ii (q \varphi j^{\, \prime})] \right] {\cal I}_{n,\ell} {\cal I}_{n-1,\ell}   
\\[3mm]
\nonumber
&& + \sqrt{\frac{2 \beta n}{q_{\mprp}^2}} \, \left [
(p \widetilde \varphi j^{\, \prime}) [(q \Lambda j) + 
\ii (q \varphi j)]  +  (p \widetilde \varphi j) 
[(q \Lambda j^{\, \prime}) - 
\ii (q \varphi j^{\, \prime})] \right] {\cal I}_{n-1,\ell-1} {\cal I}_{n,\ell-1}   
\\[3mm]
\nonumber
&& - (p \widetilde \varphi q) \left [[(j \Lambda j^{\, \prime}) - 
\ii (j \varphi j^{\, \prime})]{\cal I}_{n,\ell-1}^2 - [(j \Lambda j^{\, \prime}) +
\ii (j \varphi j^{\, \prime})]{\cal I}_{n-1,\ell}^2 \right ] 
\big \};
\eeq

%
\beq
\nonumber
&&{\cal D}^{(1)}_{AA} =  g_A g_A^{\, \prime}    
\bigg \{ [(P \widetilde \Lambda j) (p \widetilde \Lambda j^{\, \prime}) + 
(p \widetilde \Lambda j) (P \widetilde \Lambda j^{\, \prime})  - 
(j \widetilde \Lambda j^{\, \prime})(M_{\ell}^2 + m_e^2 + (p \widetilde \Lambda q))] 
\\[3mm]
\label{eq:FAA}
&& \times 
[{\cal I}_{n,\ell}^2 + {\cal I}_{n-1,\ell-1}^2]  
 + 4 \beta \sqrt{n\ell} \, (j \widetilde \Lambda j^{\, \prime}) 
{\cal I}_{n,\ell} {\cal I}_{n-1,\ell-1} 
\\[3mm]
\nonumber
&&- 
\sqrt{\frac{2 \beta \ell}{q_{\mprp}^2}} \, \left [
(P \widetilde \Lambda j) [(q \Lambda j^{\, \prime}) + 
\ii (q \varphi j^{\, \prime})]  + (P \widetilde \Lambda j^{\, \prime}) 
[(q \Lambda j) - \ii (q \varphi j)] \right] {\cal I}_{n,\ell} {\cal I}_{n,\ell-1} 
\\[3mm]
\nonumber
&& - 
 \sqrt{\frac{2 \beta \ell}{q_{\mprp}^2}} \, \left [
(P \widetilde \Lambda j)[(q \Lambda j^{\, \prime}) - 
\ii (q \varphi j^{\, \prime})]  + (P \widetilde \Lambda j^{\, \prime})  
[(q \Lambda j) + 
\ii (q \varphi j)] \right] {\cal I}_{n-1,\ell-1} {\cal I}_{n-1,\ell} 
\\[3mm]
\nonumber
&& - 
\sqrt{\frac{2 \beta n}{q_{\mprp}^2}} \, \left [
(p \widetilde \Lambda j)[(q \Lambda j^{\, \prime}) - 
\ii (q \varphi j^{\, \prime})]  +  (p \widetilde \Lambda j^{\, \prime}) 
[(q \Lambda j) + 
\ii (q \varphi j)] \right] {\cal I}_{n,\ell} {\cal I}_{n-1,\ell}   
\\[3mm]
\nonumber
&&- \sqrt{\frac{2 \beta n}{q_{\mprp}^2}} \, \left [
(p \widetilde \Lambda j) [(q \Lambda j^{\, \prime}) + 
\ii (q \varphi j^{\, \prime})]  +  (p \widetilde \Lambda j^{\, \prime}) 
[(q \Lambda j) - 
\ii (q \varphi j)] \right] {\cal I}_{n-1,\ell-1} {\cal I}_{n,\ell-1} %
\\[3mm]
\nonumber
&&+ 
 (M_{\ell}^2 + m_e^2 + (p \widetilde \Lambda q))   
 \left [[(j \Lambda j^{\, \prime}) + 
\ii (j \varphi j^{\, \prime})] {\cal I}_{n,\ell-1}^2 + [(j \Lambda j^{\, \prime}) - 
\ii (j \varphi j^{\, \prime})] {\cal I}_{n-1,\ell}^2 \right ] 
\\[3mm]
\nonumber
&&+ 
\frac{4 \beta \sqrt{n\ell}}{q_{\mprp}^2} \, 
[(q \Lambda j) (q \Lambda j^{\, \prime}) - 
(q \varphi j) (q \varphi j^{\, \prime})] {\cal I}_{n,\ell-1} {\cal I}_{n-1,\ell}
\bigg \};
\\[2mm]
\nonumber
&&{\cal D}^{(2)}_{AA} = {\cal D}^{(1)}_{AA} (q\to -q ,\, \, j \leftrightarrow j^{\, \prime}) \, .
\eeq

We notice, that the expressions  for amplitudes ${\cal M}_{VS}$, ${\cal M}_{VP}$, ${\cal M}_{VV}$ 
and ${\cal M}_{AV}$ are manifestly gauge invariant.


\end{document}